\documentclass[11pt,a4paper]{article}
\pdfoutput=1

\usepackage{jheppub}
\usepackage[T1]{fontenc}

\usepackage{mathrsfs}
\usepackage{amssymb}
\usepackage{amsmath}
\usepackage{amsthm}
\usepackage{slashed}
\usepackage{pgf}
\usepackage{pdfpages}
\usepackage{times}
\usepackage{color}
\usepackage{hyperref}
\usepackage{emptypage}
\usepackage{epstopdf}
\usepackage{pstool}

\usepackage{epsfig,latexsym,amsfonts,amsmath,amsthm,amssymb,amsbsy,multirow,slashed,color,mathrsfs,wasysym,textcomp,subfigure,wrapfig,comment}

\usepackage{color,datetime,graphicx}

\newcommand{\pink}[1]{\textcolor{\pink}{#1}}

\usepackage{chngcntr}
\counterwithout{figure}{section}
\counterwithout{figure}{subsection}

\title{Type IIA Klebanov-Strassler: the hard way}
       
\preprint{IPhT-T15/199}

\author{Giulio Pasini}

\affiliation{Institut de physique th\'eorique, Universit\'e Paris Saclay, CEA, CNRS, F-91191 Gif-sur-Yvette, France}

 \emailAdd{giulio.pasini@cea.fr}

\abstract{We construct the T-dual of the Klebanov-Strassler solution on a small region at the tip of the deformed conifold. The isometry coordinate we choose is the correct one to obtain an NS5 brane wrapping a holomorphic curve in Type IIA, as shown by a thorough analysis of the deformed conifold geometry. The shape of the locus wrapped by the NS5 brane matches the predictions from the Type IIA brane engineering construction dual to the $SU(N+M)\times SU(N)$ cascading gauge theory. The same isometry is then used to T-dualize the solution obtained by adding backreacted D3 branes to the Klebanov-Strassler solution. Our construction is the first step in a program to test the stability of antibranes in Type IIA backgrounds.}

\begin{document}
\maketitle

\section{Introduction}
Recent observations together with theoretical developments in the field of cosmology indicate that our universe has a positive cosmological constant and hence asymptotes a de Sitter space. String Theory, the only known consistent theory of quantum gravity, has numerous compactifications to four and five dimensional Anti de Sitter spaces, while no straightforward compactification to de Sitter spaces has been realized. Finding compactifications with a small positive cosmological constant is an essential task for String Theory to be a predictable theory of Physics. 

To date, the most well known procedure that uplifts AdS vacua to dS ones in String Theory is the KKLT mechanism~\cite{KKLT}. This prescribes the insertion of antibranes in long warped throats of the compactification manifold, which ensures that the uplift of the cosmological constant does not destabilize the moduli. The antibrane then completely breaks supersymmetry and its presence allows to uplift the asymptotically AdS solution to dS. Needless to say, the uplifted dS vacuum is (meta)stable only if the antibrane is (meta)stable in the original flux compactification. 

The most suitable framework where the KKLT uplift mechanism has been tested is given by the Klebanov-Strassler (KS) solution~\cite{Klebanov-Strassler}. This is a smooth four-supercharge Type IIB solution with no brane sources and only fluxes threading a nontrivial topology. In particular, the ten-dimensional spacetime is divided into a warped  four-dimensional Minkowski space and a six-dimensional internal space constituted by a deformed conifold. This is a cone over a base that topologically is equivalent to an $S_2\times S_3$, where the $S_2$ smoothly shrinks at the tip while the $S_3$ attains a finite radius and is threaded by constant fluxes. The KS solution is the model to study the validity of the KKLT uplift mechanism due to the fact that anti-D3 probe branes at the bottom of the KS throat have well-known metastable configurations~\cite{KPV}.

However, recent investigations have shown that the fate of anti-D3 branes at the bottom of the KS throat is unclear. In~\cite{Bena_existence,Bena_backreaction, Bena_bitter, Bena_metastable} their backreaction was taken into account and it was found that anti-D3 branes create a singularity in the solution. Furthermore, this singularity cannot be cloaked by a horizon~\cite{Bena_Buchel,Blaback_horizon}, which makes it problematic~\cite{gubser_good_bad_ugly}. Analogous results for antibranes in highly warped throats were found in more general contexts~\cite{Bena_M2,Blaback_smeared,Blaback_problematic, Bert-Van_Riet, Danielsson_Van-Riet}, which might lead to the conclusion that the metastability of anti-D3 branes in KS is an artifact of the probe approximation~\cite{KPV}.

On the other hand, it has been recently pointed out~\cite{Polchinski} that an analysis of the effective field theories of these probe branes should confirm the validity of their metastable configurations. 

The goal of this paper is to construct a framework to test the stability of the antibranes in warped throats for a regime of parameters that has remained unexplored so far. Specifically, our aim is to construct the T-dual version of the KS solution, which is unknown to date\footnote{Different T-dual versions are known~\cite{Minasian_hopf, Dasgupta_t-duality} for the Klebanov-Tseytlin singular theory~\cite{Klebanov-Tseytlin}, but none of these seems to be easily generalizable to KS and the results of~\cite{giapponesi} lack the mathematical rigor required for our purposes.}. The T-duality maps the radius of an $S_1$ isometry coordinate to its inverse, exchanging winding and momentum modes and giving access to the physics of a different region in the space of physical parameters. Such a new framework would then help solving the tension around the fate of antibranes in warped throats and the stability of the KKLT mechanism.

 Two different pieces of information serve as guidance to construct the T-dual version of KS. On one side, the KS solution is the gravity dual of the $\mathcal{N}=1$ cascading $SU(N+M)\times SU(N)$ gauge theory\footnote{Where $N=kM$ and $k$ is an integer.} and the corresponding Type IIA brane construction has been widely discussed in the literature~\cite{Kutasov_typeIIA_perspective, Kutasov_duality, Kutasov_branes,Dasgupta_fractional, Dasgupta_conifold,Gubser}. This construction involves $N+M$ D4 branes wrapping the four-dimensional Minkowski space and a compact direction, terminating on an NS5 brane. The latter also wraps a holomorphic curve~\cite{Kutasov_typeIIA_perspective}. 

On the other hand, the exact expression for the NS5 locus and its relation with the geometry of the deformed conifold is the second and most important clue we have. Indeed, it is general knowledge that a suitable T-duality of empty conifold geometries can give rise to solutions where an NS5 brane wraps a holomorphic locus~\cite{Bershadsky, Uranga}. Qualitatively this happens when the $U(1)$ isometry chosen for the T-duality is fibered nontrivially over a base space and has a holomorphic locus of fixed points\footnote{More specifically, the locus of fixed points should just be composed of holomorphic branches, as for the singular conifold.}. After a T-duality, this holomorphic locus will be wrapped by NS5 branes. Recently, these kinds of T-dualities have been explicitly performed on the empty geometries of the singular~\cite{royston_conifold}, resolved and deformed conifolds~\cite{royston_deformed} in a rigorous mathematical framework. In particular, as these NS5 branes arise from T-dualities of empty geometries, these techniques can still be applied to warped conifold geometries with nontrivial fluxes.

We start our analysis by showing that the isometry used in~\cite{royston_deformed} and anticipated in~\cite{Uranga} is not spoiled once one considers the full KS solution instead of the empty deformed conifold geometry. Unfortunately, even if we are granted that this isometry reproduces the desired components of the Type IIA  brane engineering construction, the parameterization of this $U(1)$ in Type IIB is extremely involved. It is possible to have a clear picture of the isometry circle only on the three-sphere at the tip of the deformed conifold and the coordinates used to write the KS solution completely hide it. 

To make progress we use the following strategy: As all the important physics to test the stability of antibranes is encoded in a region at the bottom of the KS warped throat, we focus on a small neighborhood of a particular point on the three-sphere at the tip, which we can refer to without loss of generality as the North Pole (NP). We choose this to be located on the fixed locus for our isometry. By introducing a small typical length we expand the KS solution in this neighborhood to a fixed order of precision. This allows to find a suitable set of coordinates that make the isometry of~\cite{royston_deformed} manifest. At the same time the coordinates we introduce are easily related to the global topology of the deformed conifold. We explicitly check that by T-dualizing the empty geometry we reproduce the results of~\cite{royston_deformed} expanded and written in the new NP coordinates, including the NS5 brane wrapping the desired holomorphic curve. 

We then expand and T-dualize the full KS solution in the NP neighborhood. The expansion of the fields is realized by evaluating their squares contracted with the local KS metric. Since these scalar quantities are preserved under T-duality, this ensures no loss of physically relevant information in Type IIA. As the NP is mapped on the NS5 locus in type IIA we are able to reconstruct the physics close to this region, characterized by a blowing-up dilaton and $B_2$ NS-NS field. The same technique can be carried on at an arbitrary order of precision for the expansions. 

To have a deeper insight of the KS physics and to test the stability of the antibranes, we then push our construction one step farther. We modify the KS solution by adding backreacted D3 branes localized at the NP. As this location belongs to the fixed locus of our isometry, the latter is not spoiled by the addition of the branes and thus we are able to reconstruct the corresponding T-dual solution using the same techniques as before. In particular, by expanding close to the NP, there is no need to solve the Laplace equation on the deformed conifold for the D3's, as we know that in our linearized metric the warp factor sourced by the branes will essentially coincide with that of D3 branes in flat space. In the resulting Type IIA solution we are able to see new interactions between the resulting ``background'' NS5 brane and the added D4 branes.

This paper is organized as follows. In Section~\ref{section_KS} we review the Klebanov-Strassler solution and discuss its main features. In Section~\ref{section_2} we describe the brane construction that one expects for the solution that is T-dual of KS, focusing on the NS5 locus and on the correct isometry for the T-duality. We then introduce the NP coordinates and T-dualize the local empty geometry reproducing the same results as in~\cite{royston_deformed}. In Section~\ref{section_3} we expand the full KS solution close to the NP and explicitly write down its Type IIA T-dual solution. Some consistency tests are performed on the resulting solution. In Section~\ref{section_4} we add backreacted D3 branes at the NP and apply the same procedure as before to reconstruct the corresponding T-dual solution. Section~\ref{section_discussion} is dedicated to the discussion and outlook for our future work. Many useful details and computations are relegated to the Appendix. In particular, in Appendix~\ref{appendix_coordinates} we review the geometry of the deformed conifold and introduce three coordinate systems that are widely used in our analysis. Appendix~\ref{appendix_tip} is dedicated to the study of the tip of the deformed conifold and to relating the NP coordinates to the global ones. In Appendix~\ref{appendix_isometry_proof} we show that the isometry of~\cite{royston_deformed} can be extended to the KS solution, while in Appendix~\ref{computation_expansion} we report some necessary computations to expand the KS solution in the NP neighborhood. Finally, in Appendix~\ref{section_buscher} we recall Buscher's rules for T-duality.


\section{The Klebanov-Strassler solution}\label{section_KS}
The KS solution~\cite{Klebanov-Strassler} is a Type IIB supergravity solution that preserves four supercharges and is completely smooth, as there are no localized sources but only fluxes threading a nontrivial topology. It can be thought of as the geometric transition of the singular Klebanov-Tseytlin solution~\cite{Klebanov-Tseytlin}, where D3 branes and vanishing D5 branes are placed at the tip of the (singular) conifold. The geometric transition replaces the branes with fluxes and at the same time puffs up some nontrivial cycles within the space transverse to the brane. Consequently, the overall KS metric looks like the standard D3 brane solution, with a four-dimensional warped Minkowski space and a (warped) six-dimensional internal space given by a deformed conifold. The latter is a cone over a base that is topologically equivalent to an $S_2\times S_3$. As a result of the geometric transition, the $S_3$ attains a finite radius at the tip of the cone and is threaded by constant fluxes, while the $S_2$ smoothly shrinks in this region.

To write the full KS solution it is first necessary to parameterize the deformed conifold and equip it with a Ricci-flat metric. In Appendix~\ref{appendix_coordinates} we present three different coordinate systems for this manifold, each of them suitable to show some of its specific properties. Here it suffices to say that the base of the deformed conifold is a $T^{1,1}$ space, defined in~\cite{Candelas} as the quotient manifold
\begin{equation}
T^{1,1}= \frac{SU(2)\times SU(2)}{U(1)}
\end{equation}
The $T^{1,1}$ space can be described by a combination of the Euler angles of the two $SU(2)$, consisting of two pairs of angles $ \phi_i \in [0, 2\pi[$ and $\theta_i\in [0, \pi[$ with $i=1,2$ and a coordinate $\psi = \psi_1+\psi_2 \in [0, 4\pi[$ arising from the quotient. These, together with a coordinate $\tau\geq 0$ for the radius of the cone, will be referred to as the \textit{coset coordinates} for the deformed conifold. A standard basis of one-forms was found in~\cite{Minasian}:
\begin{align}
g^1 &= \frac{e^1-e^3}{\sqrt{2}}  &  g^2 &=\frac{e^2-e^4}{\sqrt{2}} & g^3&=\frac{e^1+e^3}{\sqrt{2}} \nonumber \\
g^4&=\frac{e^2+e^4}{\sqrt{2}} &  g^5&=e^5  &  \label{KS_conifold_one_forms_1}
\end{align}
where
\begin{align}
e^1&=-\sin \theta_1\,  d\phi_1 \quad \quad e^2=d\theta_1 \quad \quad e^3=\cos \psi \sin \theta_2 \, d\phi_2-\sin \psi \, d\theta_2 \nonumber \\
e^4&=\sin \psi \sin \theta_2 \,  d\psi_2 + \cos \psi \, d\theta_2 \quad \quad 
e^5=d\psi +\cos \theta_1\,  d\phi_1 + \cos\theta_2 \, d\phi_2 \label{KS_conifold_one_forms_2}
\end{align}
Then the Ricci-flat Hyper-K\"ahler metric on the deformed conifold is~\cite{Minasian}:
\begin{equation}\label{df_metric}
ds^2_6=\frac{1}{2}\varepsilon^{\frac{4}{3}}K(\tau)\left[\frac{1}{3K^3(\tau)}[d\tau^2+(g^5)^2]+\cosh^2\frac{\tau}{2}[(g^3)^2+(g^4)^2]+\sinh^2\frac{\tau}{2}[(g^1)^2+(g^2)^2]\right]
\end{equation}
with 
\begin{equation}
K(\tau)=\frac{(\sinh 2\tau -2\tau)^{\frac{1}{3}}}{2^{\frac{1}{3}}\sinh \tau}
\end{equation}
where $\varepsilon$ is a deformation parameter. In particular, $K(\tau)$ is finite at the tip of the cone $\tau=0$ and the metric~\eqref{df_metric} becomes the metric of a three-sphere whose radius depends on $\varepsilon$, as shown in Appendix~\ref{appendix_tip}.

It is now possible to write the full KS solution. The metrics looks like the standard D3 brane ansatz:
\begin{equation}\label{KS_metric}
ds^2_{KS}=h(\tau)^{-\frac{1}{2}} \, dx^idx^i+h(\tau)^{\frac{1}{2}} ds^2_6 
\end{equation}
where $dx^i dx^i$ is the standard Minkowski metric and  $ds^2_6$ is as in~\eqref{df_metric}. The warp factor $h(\tau)$ in~\eqref{KS_metric} is given by:
\begin{equation}\label{KS_warpfactor}
h(\tau)=(g_s M \alpha^\prime)^2 \varepsilon^{-\frac{8}{3}}2^{\frac{2}{3}} I(\tau) \quad \quad \quad I(\tau)=\int_{\tau}^{\infty}dx \, \frac{x\coth x -1}{\sinh^2 x}(\sinh 2x -2x)^{\frac{1}{3}}
\end{equation}
where $I(\tau)$ in~\eqref{KS_warpfactor} attains the value $a_0\approx 0.71805$ for $\tau=0$ so that the whole spacetime~\eqref{KS_metric} is smooth. The parameter $M$ in~\eqref{KS_warpfactor} can be thought of as the quantized number of D5 charge units preserved by this solution and measured by the three-form RR field strength. The RR and NS-NS fields are written in the canonical basis of one-forms on the deformed conifold~\eqref{KS_conifold_one_forms_1}:
\begin{align}
B_2&=\frac{g_s M\alpha^\prime}{2} \, [f(\tau) g^1 \wedge g^2 +k(\tau) g^3 \wedge g^4] \label{KS_B2} \\
F_3&=\frac{M \alpha^\prime}{2}\, \{g^5\wedge g^3\wedge g^4+d[F(\tau)(g^1\wedge g^3+g^2\wedge g^4)]\} \label{KS_F3} \\
F_{5} &= \mathcal{F}_5 + \star \mathcal{F}_5 \label{KS_F5} \\
\mathcal{F}_5&=B_2\wedge F_3=\frac{g_s M^2 (\alpha^\prime)^2}{4} \, \ell(\tau)\,g^1\wedge g^2\wedge g^3\wedge g^4\wedge g^5 \nonumber \\
\star \mathcal{F}_5 &=4g_s M^2(\alpha^\prime)^2\varepsilon^{-\frac{8}{3}}\frac{\ell(\tau)}{K(\tau)^2h^2\sinh^2\tau}\,dt\wedge dx^1\wedge dx^2\wedge dx^3\wedge d\tau
\end{align}
where the functions $f,k,F,l$ are given by:
\begin{align}
f(\tau)&= \frac{\tau \coth \tau -1}{2 \sinh \tau}(\cosh\tau -1) &   k(\tau) &= \frac{\tau \coth \tau -1}{2\sinh \tau}(\cosh \tau +1) \nonumber \\
F(\tau)&=\frac{\sinh \tau -\tau}{2\sinh \tau}     &    \ell(\tau)&=\frac{\tau \coth \tau -1}{4\sinh^2\tau}(\sinh 2\tau -2\tau) \label{KS_functions}
\end{align}
The fields $B_2$ in~\eqref{KS_B2} and $F_3$ in~\eqref{KS_F3} are nonzero at the tip, while $F_5$ smoothly vanishes there. Away from the tip, $F_5$ measures $N=kM$ units of fluxes threading the $T^{1,1}$ space, where $k$ is an integer that jumps periodically with $\tau$ and is zero at the tip. Consequently, if one relates the cone coordinate $\tau$ with the energy scale of the dual gauge theory, each jump in the fluxes should represent a phase transition. Indeed, the KS solution is dual to the four dimensional $\mathcal{N}=1$  $SU(N+M)\times SU(N)$ gauge theory. Each jump in the $F_5$ flux on the $T^{1,1}$ space corresponds to a Seiberg duality~\cite{Seiberg_duality} between the $SU(N+M)\times SU(N)$ and the $SU(N+M)\times SU(N+2M)$ gauge theories. The puffing-up of the $S_3$ at the tip is due to the chiral symmetry breaking of the infrared physics of this gauge theory.


\section{T-dualizing the Klebanov-Strassler solution}\label{section_2}

\subsection{Type IIA brane constructions and the deformed conifold}\label{section_program}

The singular Klebanov-Tseytlin solution has already been T-dualized in the literature~\cite{Minasian_hopf,Dasgupta_t-duality}, but unfortunately the same techniques are not easily generalized to KS. In particular, we want to find an isometry in KS that, upon a T-duality, leads to a brane configuration that can be easily handled, as the one that realizes the gauge dual to KS in Type IIA supergravity~\cite{Kutasov_typeIIA_perspective}.

\begin{figure}
\begin{pgfpicture}
{0cm}{0cm}{10cm}{6cm}
\pgfsetxvec{\pgfpoint{0.5cm}{0cm}}
\pgfsetyvec{\pgfpoint{0cm}{0.5cm}}
\pgftranslateto{\pgfxy(7.7,6)}

\input{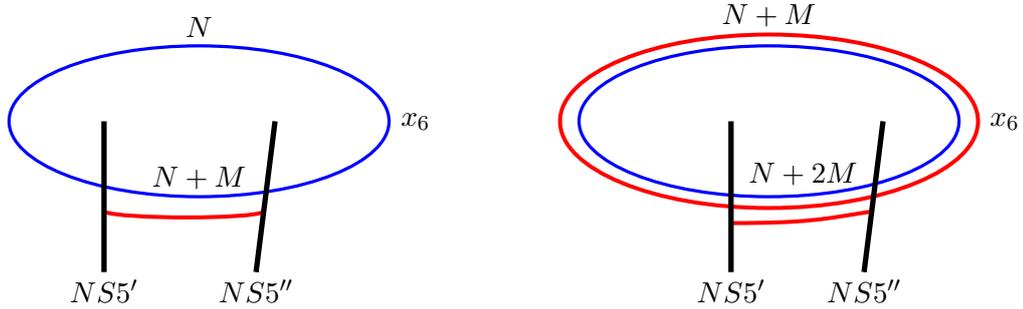}

\end{pgfpicture}
\caption{Brane configuration in the Type IIA dual to the (ultraviolet of the) $SU(N+M)\times SU(N)$ gauge theory. On the left there are $N$ D4 branes wrapping the T-duality circle $x_6$ and $M$ D4's stretched between the two NS5's for a fixed value of $x_7$. On the right the same configuration is represented for a higher value of $x_7$, where $NS5^{\prime \prime}$ has moved with respect to $NS5^\prime$ spanning a full loop around the circle and pulling the $M$ D4 branes. Now the dual theory has become $SU(N+M)\times SU(N+2M)$ and the crossing of the NS5's corresponds to the phase transition of Seiberg duality.}
\label{figure_duality}
\end{figure}

The Type IIA brane construction dual to the $\mathcal{N}=1$ $SU(N+M) \times SU(N)$ theory consists of D4 branes wrapped on a circle and intersecting the two branches of an NS5 brane. The NS5 wraps a four-dimensional Minkowski space the holomorphic locus~\cite{Witten, Kutasov_typeIIA_perspective}
\begin{equation}\label{branes_deformed_conifold}
z_1z_2=-\frac{\varepsilon^2}{2} 
\end{equation}
where we have defined $z_1=x_4+ix_5$ and $z_2=x_8+ix_9$. Away from the origin, namely for $|z_1|,|z_2|>>0$, one can approximate $\varepsilon\simeq 0$ in~\eqref{branes_deformed_conifold} and hence consider the NS5 as two separate branes,  the first one wrapping the directions $x_4$ and $x_5$ and located at $x_8=x_9 =0$ and vice-versa for the other. The coordinate $x_6$ parameterizes a circle wrapped by $N+M$ D4 branes, that also wrap the four-dimensional Minkowski space. The two NS5's intersect the compact direction $x_6$ in two distinct points. Among the D4 branes, only $N$ of them wrap the whole $x_6$, while the remaining $M$ wrap the same interval on $x_6$ delimited by the NS5 branes and terminate on them. This setup gives rise to a dual $\mathcal{N}=1$ $SU(N+M) \times SU(N)$ gauge theory. Furthermore, the intersections of the two NS5 branches on the circle $x_6$ depend on the remaining noncompact coordinate $x_7$: indeed the NS5 branches bend in this direction pulled by the D4. Therefore, while there are always $M$ D4's between the two NS5's, the number of D4's wrapping the whole $x_6$ depends on how many times the NS5's have spiraled around it and hence depends on $x_7$. From the point of view of the dual theory the spiraling of the NS5's causes a cascade of Seiberg dualities~\cite{Kutasov_duality} and $x_7$ qualitatively plays the role of the energy scale of the gauge theory. This mechanism along with the Type IIA brane construction is represented in Figure~\ref{figure_duality}.

While the picture described so far is accurate in the ultraviolet where $|z_1|,|z_2|>>1$ in~\eqref{branes_deformed_conifold}, this is not so in the infrared, namely close to the origin of the coordinates. The chiral symmetry breaking that takes place in the infrared of the $SU(N+M) \times SU(N)$ theory is paralleled in the gravity dual by the joining of the two NS5 branches into a single holomorphic curve via the parameter $\varepsilon$. The smoothing of the brane locus due to quantum infrared physics is reminiscent of the geometric transition from the Klebanov-Tseytlin to the Klebanov-Strassler solution and as we will see this is not a coincidence.

The holomorphic locus expected for the NS5~\eqref{branes_deformed_conifold} uniquely fixes the isometry that one has to use to T-dualize the KS solution, which is a common feature for conifold geometries.
It is well known in the literature that only a suitable T-duality of empty conifold geometries can lead to NS5 brane configurations~\cite{Uranga, Bershadsky}. Schematically, this is made possible because the conifold geometries can be seen as nontrivial fibrations whose base spaces contains smooth loci where the fiber degenerates. Thus, choosing a $U(1)$ isometry on the fiber, the T-duality circle smoothly shrinks at the degeneration locus and hence blows up there in the T-dual solution, as is clear from Buscher's rules - see Appendix~\ref{section_buscher}. The T-duality hence gives a metric, dilaton and $B_2$ fields that blow up with the appropriate power on these loci, and $H_3=dB_2$ measures an integer NS5 charge. This was rigorously shown for the empty singular conifold~\cite{royston_conifold} and for the empty resolved and deformed conifolds\footnote{By empty deformed conifold here we mean a supergravity solution given by just the metric $ ds^2=dx^i dx^i + ds^2_6$ where $ds^2_6$ is as in \eqref{df_metric}. This is a valid supergravity solution as the deformed conifold is Ricci-flat.} in~\cite{royston_deformed}. In particular, the authors of~\cite{royston_deformed} managed to get an NS5 wrapping precisely the locus~\eqref{branes_deformed_conifold}, and the procedure requires some knowledge of the properties of the deformed conifold reviewed in Appendix~\ref{appendix_coordinates}. Here it suffices to say that deformed conifold is a six dimensional manifold embedded in $\mathbb{C}^4$ via
\begin{equation}\label{conifold_brane_coordinates}
z_1z_2-xu=-\frac{\varepsilon^2}{2} 
\end{equation}
where $z_1, z_2, x, u \, \in \mathbb{C}$ subject to~\eqref{conifold_brane_coordinates} will be called the \textit{brane coordinates} for the deformed conifold. The choice for the notation of~\eqref{conifold_brane_coordinates} will be related to that of~\eqref{branes_deformed_conifold} in a moment. The brane coordinates describe the deformed conifold as a fibration, where the base space is a $\mathbb{C}^2$ parametrized by $z_1,z_2$, while the fiber is parametrized by either $x$ or $u$. Indeed, two charts $\mathcal{U}_1=\{x\neq 0\}$ and $\mathcal{U}_2=\{u \neq 0 \}$ are needed to cover the deformed conifold: in the following we will always assume that $x\neq 0$. The whole discussion can be rewritten for the other chart by simply replacing $x\leftrightarrow u$. The relationship between the brane coordinates and the coset ones of Section~\ref{section_KS} is presented in~\eqref{branes_to_coset}.
For the deformed conifold note the existence of a $U(1)$ on the fiber of~\eqref{conifold_brane_coordinates} that acts as 
\begin{align}
x \rightarrow e^{i\xi} x  \quad  \quad u\rightarrow e^{-i\xi} u \quad \quad \xi \in \mathbb{R} \label{isometry}
\end{align}
Equation \eqref{isometry} is a symmetry for the conifold as written in \eqref{conifold_brane_coordinates} and it was proved in~\cite{royston_deformed} that this an isometry for the conifold metric~\eqref{df_metric}. The locus that is left invariant by~\eqref{isometry} coincides precisely with~\eqref{branes_deformed_conifold} an gets wrapped by an NS5 in the T-dual solution.  In~\cite{royston_deformed} this was also confirmed by computing the integer NS5 brane charge with the NS-NS three-form field strength that one gets in Type IIA.

Incidentally, note that taking $\varepsilon =0$ in \eqref{conifold_brane_coordinates} one obtains the defining equation for the singular conifold, which is the internal manifold of the Klebanov-Tseytlin solution~\cite{Klebanov-Tseytlin}. Then T-dualizing along the same $U(1)$ as in~\eqref{isometry} one ends up with two NS5 branes wrapping~\eqref{branes_deformed_conifold} with $\varepsilon =0$. The Klebanov-Tseytlin and KS solutions essentially coincide in the ultraviolet and this happens also for the NS5 brane loci that one gets from a T-duality of their geometries. These two theories however differ in the infrared, where the Klebanov-Tseytlin solution is singular and the related NS5 locus remains composed by two separate branches. In KS, the chiral symmetry breaking is responsible for the puffing-up of the $S_3$ at the tip, which in the T-dual solution is paralleled by the joining of the two NS5 branches into a single holomorphic curve. 

The isometry~\eqref{isometry} is the right one to obtain the NS5 configuration~\eqref{branes_deformed_conifold} expected from the brane construction dual to the $SU(N+M)\times SU(N)$ theory, starting from the empty deformed conifold.  A priori it is not obvious that~\eqref{isometry} remains an isometry for the full KS solution, but if it is we already know that a T-duality along it would give the NS5 configuration~\eqref{branes_deformed_conifold} that is required from the brane construction. Indeed, this NS5 configuration is a specific feature of the geometry of the deformed conifold itself, which is preserved if one equips the deformed conifold with additional warping and fields as in the KS solution. 

Happily enough, we prove that~\eqref{isometry} is an isometry even for the full KS solution. In Appendix~\ref{appendix_isometry_proof} we show that~\eqref{isometry} is a particular transformation of an $SO(4)$ group that leaves the deformed conifold~\eqref{conifold_brane_coordinates} invariant. We then show that the coordinate $\tau$ is not affected by the $SO(4)$ transformations and therefore all the KS functions in~\eqref{KS_functions} are left invariant together with the warp factor~\eqref{KS_warpfactor}. Secondly, as the fields $B_2$ in~\eqref{KS_B2} and $F_3$ in~\eqref{KS_F3} can be rewritten in an $SO(4)$-invariant form~\cite{remarks}, one concludes that~\eqref{isometry} is an isometry for the whole KS background.

We have understood how to T-dualize the KS solution to a Type IIA one similar to that depicted in Figure~\ref{figure_duality}, expected from the dual gauge theory. To be precise, a T-duality along~\eqref{isometry} will result in a setup similar to that in Figure~\ref{figure_duality}, but smeared along the T-duality circle, the equivalent of $x_6$ in the Figure.  We will discuss in the next section how to perform this T-duality.


\subsection{The North Pole expansion}\label{section_practice}

In the previous Section we found the isometry that allows to T-dualize the KS solution and obtain a similar brane configuration expected from the dual gauge theory,  smeared along the T-duality direction. However, putting in practice this strategy proves to be most tricky. First of all, making the isometry~\eqref{isometry} manifest in the KS solution requires a laborious change of coordinates on the deformed conifold. The transformation~\eqref{isometry} can be identified with a shift in the complex phase of the brane coordinate $x$, when $x\neq 0$. In Appendix~\ref{appendix_coordinates}, equation~\eqref{branes_to_coset} shows the relationship between the brane coordinates of~\eqref{conifold_brane_coordinates} and the coset coordinates used to write the KS solution in Section~\ref{section_KS}. The phase of $x$ written in coset coordinates is a highly nontrivial and ill-defined function. This is because one also has to deal with the $x=0$ locus in~\eqref{conifold_brane_coordinates}, where the phase of $u$ in~\eqref{isometry} becomes the valid coordinate for the isometry instead. Secondly, it is quite hard to visualize the isometry circle~\eqref{isometry} on the deformed conifold and we only have a clear picture of it on the $S_3$ at the tip - see Appendix~\ref{appendix_tip}. While the authors of~\cite{royston_deformed} took care of these subtleties for the empty deformed conifold, this ends up in untreatable formulas if performed on the full KS solution that can hide the interesting physics encoded in Type IIA. 

These difficulties are encountered if one attempts to T-dualize the KS solution as a whole, but might well be avoided if one compromises to reconstruct the T-dual solution of just a small region of the deformed conifold in the KS solution. This certainly does not invalidate the possibility to check the antibrane stability in Type IIA. Indeed, the stability of the antibrane should be checked close to the (image of the) tip of the deformed conifold and hence it suffices to choose a small region there. 

We intend to realize the program of Section~\ref{section_program} in the following way. We focus on a small region on the deformed conifold, requiring it to be a small neighborhood centered on a point on the $S_3$ at the tip that will be called the North Pole (NP). This Section is dedicated to the construction of such neighborhood and to finding a good set of coordinates so that the metric~\eqref{df_metric} gets linearized to an $\mathbb{R}^6$ metric around the NP and the isometry~\eqref{isometry} becomes manifest. We then T-dualize just the empty geometry of the small neighborhood using the isometry~\eqref{isometry} and show that we get an NS5 brane wrapping the curve~\eqref{branes_deformed_conifold}. Our local results are then compared with the globally-valid ones of~\cite{royston_deformed}. In the next sections we expand the KS solution in the NP neighborhood and then T-dualize it to realize the program of Section~\ref{section_program}. 

To construct the NP neighborhood we introduce a small parameter $\delta$ that will be used to linearize the KS background. We then define a new coordinate system suitable to both linearize the metric~\eqref{df_metric} and to make the isometry~\eqref{isometry} manifest, which is performed in two steps. First of all, we redefine the coset coordinates of the deformed conifold in~\eqref{df_metric}, constraining some combinations of them to be of order $\delta$~\cite{Bena_giant}:\begin{align}
\alpha &= \frac{\theta_1+\theta_2}{2}  &  \beta&=\frac{\phi_1+\phi_2}{2} &  \tilde{\tau}&=\frac{\tau}{2}\delta \nonumber \\
 \omega& =\frac{\phi_2-\phi_1}{2}\delta &   \nu &=\left(\frac{\theta_1-\theta_2}{2}-\frac{\pi}{2}\right)\delta  &  \mu&=\frac{\pi-\psi}{2}\delta \label{np_expansion}
\end{align}
The tilde from $\tilde{\tau}$ will always be dropped, keeping in mind the rescaling of a factor of two. Secondly, $\omega, \nu, \mu$ in~\eqref{np_expansion} are replaced with the following combinations:
\begin{align}
r&=\sqrt{\mu^2\cos^2 \alpha +\nu^2} \nonumber \\
z&=\omega -\mu \sin \alpha \nonumber \\
\sigma &=\arctan\left[\frac{-\nu}{\mu \cos \alpha}\right] \label{spherical_cylindrical}
\end{align}
The coordinates $\tau, \alpha, \beta, r, z, \sigma$ in~\eqref{np_expansion} and~\eqref{spherical_cylindrical} will be referred to as the NP coordinates. Note that these are composed by three angular coordinates $\alpha, \beta ,\sigma$ and three radial coordinates of order $\delta$, namely $\tau, r,z$. The NP is located at $\tau=r=z=0$ and hence lies on the $S_3$ at the tip of the conifold. To avoid clutter, the parameter $\delta$ will be suppressed in the formulas where it is not necessary, keeping in mind that only $\tau, r, z$ carry a power of $\delta$. In addition, the base one-forms $d\tau, dr, dz$ will be considered of order $\delta$, meaning that once one expresses these as functions of the coset coordinates they get multiplied by a factor of $\delta$. 

The conifold metric~\eqref{df_metric} expanded to lowest order in the NP coordinates then becomes:
\begin{equation}\label{metric_NP}
ds^2_6 \simeq \varepsilon^{\frac{4}{3}}\left(\frac{2}{3}\right)^{\frac{1}{3}}[d\tau^2+\tau^2(d\alpha^2+\cos^2\alpha \,d\beta^2)+dr^2+dz^2+r^2\, (d\sigma+ d\beta)^2]
\end{equation}
Note that while each term in~\eqref{metric_NP} is of order $\delta^2$, the contraction of~\eqref{metric_NP} with itself gives a scalar of order one. The metric~\eqref{metric_NP} is not quite the metric of an $\mathbb{R}^6$, which can be recovered by adding $\beta$ to the definition of $\sigma$ in~\eqref{spherical_cylindrical}. The reason why it is necessary to write the metric~\eqref{metric_NP} keeping the cross-term $d\sigma+d\beta$ is related to the shape of the NS5 locus in the NP neighborhood and will become clear in a moment. 

The NP neighborhood is composed by two three-dimensional subspaces. The first one (spanned by $\tau$, $\alpha$ and $\beta$) is written in spherical coordinates, while the second one is parameterized by cylindrical coordinates. These two subspaces have a direct connection with the topology of the deformed conifold. The two sphere that shrinks at $\tau=0$ in~\eqref{metric_NP} is \textit{exactly} the $S_2$ that shrinks at the tip of the conifold, while the remaining subspace parameterizes the portion of the NP neighborhood that lies on the $S_3$. A detailed interpretation for the coordinates $r,\sigma, z$ and their connection with the $S_3$ is reported in Appendix~\ref{appendix_tip}. 

To show that the NP expansion makes it easier to realize the program of Section~\eqref{section_program} it is useful to T-dualize the empty metric~\eqref{metric_NP}. Indeed~\eqref{metric_NP} is the linearized metric on a small neighborhood of the empty deformed conifold and one can compare the local physics it exhibits with the globally-valid T-duality of~\cite{royston_deformed}. The first step is to expand the brane coordinates of~\eqref{conifold_brane_coordinates} using the NP ones. Equation~\eqref{branes_to_coset} in Appendix~\ref{appendix_coordinates} reports the coordinate change between brane and coset coordinate systems. Inserting~\eqref{np_expansion} and~\eqref{spherical_cylindrical} into~\eqref{branes_to_coset} and expanding to lowest order one gets:
\begin{align}
x&\simeq \frac{\varepsilon}{\sqrt{2}}[r\cos \sigma+i(\tau \cos \alpha +r\sin \sigma)]e^{i\beta} \nonumber \\
u&\simeq \frac{\varepsilon}{\sqrt{2}}[r\cos \sigma+i(\tau \cos \alpha -r\sin \sigma)  ]e^{-i\beta} \nonumber \\
z_1&\simeq \frac{\varepsilon}{\sqrt{2}}[z+i(1-\tau \sin \alpha)] \nonumber \\
z_2&\simeq \frac{\varepsilon}{\sqrt{2}}[-z +i(1+\tau \sin \alpha)] \label{brane_coordinates_np}
\end{align}
One can observe two crucial facts from~\eqref{brane_coordinates_np}. First of all, $x$ and $u$ in~\eqref{brane_coordinates_np} are of order $\delta$, while $z_1$ and $z_2$ are of order one with corrections of order $\delta$. The NP, that corresponds to $\tau=r=z=0$, is precisely on the locus~\eqref{branes_deformed_conifold}. Secondly, comparing~\eqref{isometry} and~\eqref{brane_coordinates_np} it is clear that $\beta$ becomes the coordinate that parameterizes the T-duality circle, as it is a full angular coordinate in the definition~\eqref{np_expansion}. 

\begin{figure}
\begin{pgfpicture}
{0cm}{0cm}{10cm}{5.5cm}
\pgfsetxvec{\pgfpoint{0.5cm}{0cm}}
\pgfsetyvec{\pgfpoint{0cm}{0.5cm}}
\pgftranslateto{\pgfxy(17,10)}

\input{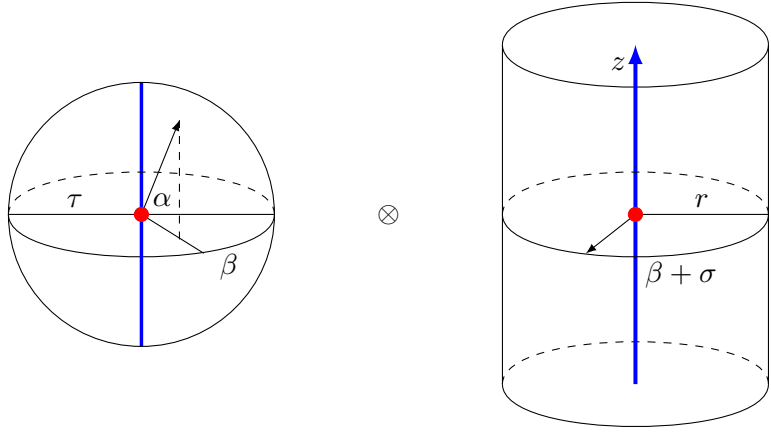}

\end{pgfpicture}
\caption{The NP neighborhood seen as a product of the collapsing $S_2$ at $\tau=0$ and a cylinder that entirely lies on the $S_3$. The red dots represent the NP, while the blue lines represent the NS5 brane locus~\eqref{brane_locus_np_2}}
\label{figure_NP}
\end{figure}

We now rewrite the locus~\eqref{branes_deformed_conifold} in the NP coordinates. Note that inserting~\eqref{brane_coordinates_np} into~\eqref{conifold_brane_coordinates} one no longer obtains an equality, as~\eqref{brane_coordinates_np} was obtained expanding to highest order in $\delta$. Consequently, as~\eqref{branes_deformed_conifold} is satisfied at lowest order in the NP coordinates, to find the brane locus one has to also impose $x=u=0$ to hold at lowest order as well. From~\eqref{brane_coordinates_np} we get
\begin{equation}\label{brane_locus_np_2}
x=u=0 \quad \Longleftrightarrow \quad \cos \alpha=r=0
\end{equation}
which is the locus that gets wrapped by an NS5 in Type IIA. The neighborhood around the NP parameterized by the coordinates used in~\eqref{metric_NP} together with the NS5 locus is represented in Figure~\ref{figure_NP}.  

The T-duality along $\beta$ of the empty geometry~\eqref{metric_NP} confirms that~\eqref{brane_locus_np_2} gets wrapped by an NS5 brane in Type IIA. Following Buscher's rules  reported in Appendix~\ref{section_buscher}, the metric~\eqref{metric_NP} should be rewritten as:
\begin{align}
ds^2_6 =& \, \varepsilon^{\frac{4}{3}}\left(\frac{2}{3}\right)^{\frac{1}{3}}\left[(\tau^2\cos^2\alpha +r^2)\left(d\beta +\frac{r^2}{\tau^2\cos^2\alpha +r^2} \,d\sigma\right)^2+d\tau^2+\tau^2 \, d\alpha^2+dr^2+dz^2 \nonumber \right.\\
  &+ \left. \frac{r^2\tau^2\cos^2\alpha}{\tau^2\cos^2\alpha+r^2}\,d\sigma^2\right] \label{metric_NP_tduality}
\end{align}
and we define the quantity:
\begin{equation}\label{amu}
A_{\sigma}\, d\sigma=\frac{r^2}{\tau^2\cos^2\alpha +r^2} \, d\sigma
\end{equation}
Then a T-duality along $\beta$ maps the NP neighborhood to a region in Type IIA, where the local metric is given by:
\begin{align}
ds^2_{6, \,IIA} =& \, \varepsilon^{\frac{4}{3}}\left(\frac{2}{3}\right)^{\frac{1}{3}}\left(d\tau^2+\tau^2 \, d\alpha^2+dr^2+dz^2+\frac{r^2\tau^2\cos^2\alpha}{\tau^2\cos^2\alpha+r^2}\,d\sigma^2\right) \nonumber \\
&+\varepsilon^{-\frac{4}{3}}\left(\frac{2}{3}\right)^{-\frac{1}{3}}\frac{d\beta^2}{\tau^2\cos^2\alpha +r^2} \label{metric_NP_IIA}
\end{align}
In addition, in Type IIA one gets a nontrivial dilaton:
\begin{equation}\label{IIa_dilaton}
e^{2\Phi}= \varepsilon^{-\frac{4}{3}}\left(\frac{2}{3}\right)^{-\frac{1}{3}}\frac{1}{\tau^2 \cos^2 \alpha +r^2}
\end{equation}
and a nontrivial $B_2$ field:
\begin{equation}\label{IIa_b2}
B_2 = \frac{r^2}{\tau^2 \cos^2 \alpha+ r^2} \, d\sigma \wedge d\beta 
\end{equation}
The $g_{\beta \beta}$ component of~\eqref{metric_NP_IIA}, the dilaton~\eqref{IIa_dilaton} and $B_2$ in~\eqref{IIa_b2} blow up precisely on the locus~\eqref{brane_locus_np_2} as one would expect in a solution containing NS5 branes. We have also verified that~\eqref{metric_NP_IIA}, \eqref{IIa_dilaton} and~\eqref{IIa_b2} represent the NP expansion of the corresponding quantities found in the analogous T-duality of the empty deformed conifold in~\cite{royston_deformed}. In~\cite{royston_deformed} it was also shown that the flux of $H_3=dB_2$ found in Type IIA measures an integer NS5 charge. Qualitatively, this is also confirmed by the fact that $B_2$ in~\eqref{IIa_b2} does not contain any factor of $\epsilon$, which is the only physically relevant constant of the empty geometry. It is important to stress that the shape of the NS5 brane in Type IIA is a feature of the T-duality along the particular isometry we chose and the manifold we are working with. Both these ingredients are still there in the KS solution and hence we expect also the same NS5 to appear in the T-dual version of this solution. 

If~\eqref{metric_NP} had been written exactly as an $\mathbb{R}^6$ metric the situation would be radically different. Indeed, a T-duality along $\beta$  would not produce any $B_2$ field, while $g_{\beta \beta}$ and the dilaton in Type IIA would blow up on a locus that is different from~\eqref{brane_locus_np_2}, which verifies $u=x=0$ in NP coordinates. The reason why the NP metric should be written as in~\eqref{metric_NP} lies in the fact that the T-duality circle as defined in~\eqref{isometry} wraps both the shrinking $S_2$ at the tip and the blown up $S_3$, as explained in Appendix~\ref{appendix_tip}.


\section{The Type IIA solution T-dual to KS}\label{section_3}

\subsection{Expansion of the KS solution in the NP neighborhood}\label{section_expansion}
In this section we expand the KS solution around the NP and rewrite it in the formalism of Buscher's rules of Appendix~\ref{section_buscher} to ease the T-duality in $\beta$. 

We expand the KS background of Section~\ref{section_KS} starting from the metric~\eqref{KS_metric}. The deformed conifold metric~\eqref{df_metric} becomes as in~\eqref{metric_NP} and is of order $\delta^2$, while the Minkowski metric $dx^idx^i$ on the first four coordinates remains untouched.
Requiring to keep corrections up to order $\delta^2$ in the metric determines how to expand the warp factor~\eqref{KS_warpfactor}. In particular, one needs to truncate differently the expansions of $h^{-\frac{1}{2}}$ and $h^{\frac{1}{2}}$, which are then denoted with a hat:
\begin{align}
\hat{h}^{-\frac{1}{2}} &= \frac{\varepsilon^{\frac{4}{3}}}{g_sM\alpha^\prime 2^{\frac{1}{3}}}\left(\frac{1}{\sqrt{a_0}}-\frac{a_2 \tau^2}{2a_0^{\frac{3}{2}}}\right) \nonumber \\
\hat{h}^{\frac{1}{2}} & = g_sM\alpha^\prime \varepsilon^{-\frac{4}{3}}2^{\frac{1}{3}}\sqrt{a_0} \label{warpfactor_NP}
\end{align}
where $a_0\approx 0.71805$ was computed in~\cite{Klebanov-Strassler} and $a_2=-2^{\frac{8}{3}}\cdotp 3^{-\frac{4}{3}}$ is computed in Appendix~\ref{computation_expansion}. Note that in~\eqref{warpfactor_NP} $\hat{h}^{-\frac{1}{2}}$ was truncated at order $\delta^2$, while  $\hat{h}^{\frac{1}{2}}$ was truncated at order one, so that all the metric components contribute with terms up to order two. Indeed, the full expanded KS metric becomes
\begin{equation}\label{metric_expanded}
ds^2_{KS}=\hat{h}^{-\frac{1}{2}} dx^i dx^i + \hat{h}^{\frac{1}{2}}ds^2_6
\end{equation}
with $ds^2_6$ as in~\eqref{metric_NP} and the warp factor as in~\eqref{warpfactor_NP}. To perform a T-duality along $\beta$ one rewrites $ds^2_6$ in~\eqref{metric_expanded} exactly as in~\eqref{metric_NP_tduality} and defines the same quantity $A_\sigma d\sigma$ as in~\eqref{amu}.

Now that the metric has been expanded one can proceed with the expansion of the KS RR and NS-NS field strengths. It is not possible to simply expand a field strength in power series and then just truncate it at some fixed order in $\delta$, as this might lead to a loss of physically relevant information. To this purpose, we use a more reliable procedure that consists of two steps. Given an $n$-form field strength $F_{\mu_1 \mu_2...\mu_n}$ one computes its square $(F_n)^2$ defined as
\begin{equation}\label{rule_expansion}
(F_n)^2 = F_{\mu_1 \mu_2...\mu_n}g^{\mu_1 \nu_1}g^{\mu_2 \nu_2}\cdotp \cdotp \cdotp g^{\mu_n \nu_n}F_{\nu_1 \nu_2...\nu_n} 
\end{equation}
where in~\eqref{rule_expansion} one has to use the expanded metric~\eqref{metric_expanded}. Then one first truncates the power series expansion in $\delta$ of $(F_n)^2$ at a fixed order. Secondly, one expands $F_{\mu_1 \mu_2...\mu_n}$ in power series and keeps only the terms that contribute to the truncation of $(F_n)^2$. This criterion is mathematically accurate, as it based on the expansion of~\eqref{rule_expansion}, which is a scalar. Most importantly, this criterion is also physically meaningful. The square of a field strength~\eqref{rule_expansion} is of the same order as (the square of) the flux that the field is carrying and this guarantees no loss of relevant information. Furthermore, the scalars built as in~\eqref{rule_expansion} are preserved under a T-duality together with their power series expansions. This means that if one expands the field strengths in Type IIB following the procedure described above then in Type IIA one automatically reconstructs the field strengths expanded with the very same criterion. Using this rule we can safely proceed to rewrite the KS fields in the NP coordinates of Section~\ref{section_practice}. We choose to keep all the terms in the expansions of the KS field strengths that contribute to the lowest order term of the expansion of their square.

The expansion of the $B_2$ KS field in~\eqref{KS_B2} rewritten directly in the formalism of Buscher's rules is given by
\begin{equation}
B_2=B_{a \beta}\, dy^{a}\wedge(d\beta +A_\sigma \, d\sigma)+\widehat{B}_2
\end{equation}
where $A_\sigma \, d\sigma$ is as in~\eqref{amu} and 
\begin{align}
B_{a \beta}\, dy^a =& \frac{2}{3}M g_s\alpha^\prime \tau(\tau^2 \cos\alpha\, d\alpha -r\sin \alpha \, dr-r\cos \sigma \cos \alpha \, dz) \label{bibeta} \\
\widehat{B}_2 =&-\frac{2}{3}M g_s\alpha^\prime \tau(r\cos\alpha \cos \sigma dz\wedge d\sigma +\cos\alpha \sin \sigma dz\wedge dr+r\sin \alpha dr\wedge d\sigma)\nonumber \\
& -B_{a \beta}dy^a\wedge A_\sigma \, d\sigma \label{B2_NP}
\end{align}
It is useful to show how to count the factors of $\delta$ in $B_2$ and its square. According to the conventions of Section~\ref{section_practice}, the coordinates $\tau, r, z$ along with $d\tau, dr, dz$ carry a factor of $\delta$: consequently, all the terms appearing in~\eqref{B2_NP} are of order $\delta^3$. As $B_2$ has legs only along the deformed conifold, when one builds its square as in~\eqref{rule_expansion} one has to use the inverse expanded metric along the deformed conifold, which is of order $\delta^{-2}$. Therefore $(B_2)^2$ is of order $\delta^2$, but the physically meaningful information is carried by $H_3=dB_2$, whose square is of order one. Indeed, taking the differential of $B_2$ in~\eqref{B2_NP} does not alter the order of magnitude of the single terms, which remains $\delta^3$. Now $H_3$ has one more leg along the deformed conifold with respect to $B_2$ and hence $(H_3)^2$ receives an additional factor $\delta^{-2}$ from the expansion of the inverse metric on the conifold. The fact that $H_3$ is not irrelevant at the NP is expected from the discussion of the KS solution in Section~\ref{section_KS}: as the flux of $H_3$ on the $S_3$ remains finite even at the tip then this form should be of order one close to the NP.

Using the conventions of Appendix \ref{section_buscher} we rewrite the expansion of $F_3$ in~\eqref{KS_F3} as
\begin{equation}
F_3 = F_{3, \beta}\wedge(d\beta + A_\sigma \, d\sigma)+\widehat{F}_3
\end{equation}
where $A_\sigma d\sigma$ is as in \eqref{amu}. Then we find:
\begin{align}
F_{3,\beta} =& \,M\alpha^{\prime} \left(-\frac{1}{3}r \tau \cos \sigma d\tau \wedge d\alpha+\frac{1}{3}\tau \cos \alpha \cos \sigma \sin \alpha d\tau \wedge dr + \frac{1}{3}\tau \cos^2\alpha d\tau \wedge dz \right. \nonumber \\
 &  \left. -\frac{1}{3}r\tau \cos \alpha \sin \alpha \sin \sigma d\tau \wedge d\sigma +\frac{2}{3}\tau^2 \cos^2\alpha \cos \sigma d\alpha \wedge dr \right. \nonumber \\
 & \left. -\frac{2}{3}\tau^2\cos \alpha \sin \alpha d\alpha \wedge dz -\frac{2}{3}r\tau^2 \cos^2\alpha \sin \sigma d\alpha \wedge d\sigma + 2r dr \wedge dz \right) \label{f3b}
 \end{align}
and 
\begin{align}
\widehat{F}_3 =& -\frac{1}{3}M\alpha^\prime\tau \sin \sigma d\tau \wedge d\alpha \wedge dr -M\alpha^{\prime}\frac{r\tau \cos \alpha}{3(r^2+\tau^2\cos^2\alpha)}\left(\tau^2\cos\alpha \cos \sigma d\tau\wedge d\alpha \wedge d\sigma \right. \nonumber \\
& \left. +r\cos \sigma \sin \alpha d\tau \wedge dr \wedge d\sigma +r\cos \alpha d\tau \wedge dz \wedge d\sigma +2r\tau\cos \alpha \cos \sigma d\alpha\wedge dr \wedge d\sigma \right. \nonumber \\
& \left. + 6 r \tau \sin \alpha d\alpha \wedge dz \wedge d\sigma \right) \label{hatf3}
\end{align}
The expansions of some wedge products among the base one-forms~\eqref{KS_conifold_one_forms_1}  are reported in Appendix~\ref{computation_expansion}. Each term of~\eqref{f3b} and~\eqref{hatf3} carries a factor $\delta^3$, which implies that $F_3^2$ is of order one. This is again expected from the discussion at the end of Section~\ref{section_KS}: as $F_3$ is constant and nonzero at the tip its square has to be of order one in the NP neighborhood.

The NP expansion of the self-dual RR five-form~\eqref{KS_F5} is rewritten as
\begin{equation}
F_5 =F_{5,\beta} \wedge (d\beta +A_\sigma \, d\sigma)+\widehat{F}_5
\end{equation}
where 
\begin{align}
F_{5,\beta}&=-g_s M^2(\alpha^\prime)^2\frac{16\tau^3 r}{9}\,\cos\alpha \, d\alpha \wedge dr\wedge dz\wedge d\sigma \nonumber \\
\widehat{F}_5&=\frac{\varepsilon^{\frac{8}{3}}}{g_s^3M^2(\alpha^\prime)^2a_0^2}\frac{4}{3^{\frac{4}{3}}}\tau\, dt\wedge dx^1\wedge dx^2\wedge dx^3\wedge d\tau \label{F5_NP}
\end{align}
We stress that $\widehat{F}_5$ is just the expansion of the original $\star F_5$ in~\eqref{KS_F5}, as $F_{5,\beta}\wedge A_\sigma \, d\sigma =0$. $(F_5)^2$ is of order $\delta^2$ and then its flux is small. This is also expected from the physics of the KS solution: indeed the flux measured by $F_5$ on the $T^{1,1}$ space depends on $\tau$ and smoothly goes to zero at the tip.


\subsection{The Type IIA solution dual to KS} \label{section_IIA_KS}
The T-dual of the KS solution in the NP neighborhood is readily obtained from the results of Section~\ref{section_expansion}. Using the expansions for the warp factor~\ref{warpfactor_NP} the full Type IIA metric at the NP is given by:
\begin{align}
ds^2_{IIA}=\, &\frac{\varepsilon^{\frac{4}{3}}}{g_sM\alpha^\prime 2^{\frac{1}{3}}}\left(\frac{1}{\sqrt{a_0}}-\frac{a_2 \tau^2 }{2a_0^{\frac{3}{2}}}\right)\, dx_i dx_i+\frac{3^{\frac{1}{3}}d\beta^2}{g_sM\alpha^\prime 2^{\frac{2}{3}}\sqrt{a_0}(\tau^2\cos^2\alpha +r^2)} \nonumber \\
&+\,\frac{2^{\frac{4}{3}} 3^{\frac{1}{3}}\tau}{3\sqrt{a_0}(\tau^2\cos^2\alpha +r^2)}\, d\beta \,(\tau^2 \cos\alpha\, d\alpha -r\sin \alpha \, dr-r\cos \sigma \cos \alpha \, dz)\nonumber \\
&+\, g_sM\alpha^\prime \sqrt{a_0}\frac{2^{\frac{2}{3}}}{3^{\frac{1}{3}}}\left(d\tau^2+\tau^2 d\alpha^2+dr^2+dz^2+\frac{r^2\tau^2\cos^2\alpha}{\tau^2\cos^2\alpha+r^2}d\sigma^2\right) \label{metric_IIA}
\end{align}
while the Type IIA dilaton is nontrivial:
\begin{equation} \label{dilaton_IIA}
e^{2\Phi}=\frac{3^{\frac{1}{3}}}{g_sM\alpha^\prime 2^{\frac{2}{3}}\sqrt{a_0}(\tau^2\cos^2\alpha +r^2)}
\end{equation}
The first and third lines in~\eqref{metric_IIA} together with the dilaton~\eqref{dilaton_IIA} are similar to  the corresponding lines in~\eqref{metric_NP_IIA} and the dilaton~\eqref{IIa_dilaton} that one gets by T-dualizing the empty geometry expanded around the NP. The fact that now we are T-dualizing the KS solution is signaled by the presence of the KS constants such as $M$ and $a_0$ and by the $\tau^2$-correction in the metric on the Minkowski space. The second line of~\eqref{metric_IIA} is completely new and contains cross-terms with $\beta$ entirely coming from the nontrivial expansion of the KS $B_2$ field in~\eqref{B2_NP}.  From \eqref{metric_IIA} one can easily verify that $g_{\mu \nu}g^{\mu \nu}$ is still of order one,  as expected. Note how the $g_{\beta \beta}$ component in~\eqref{metric_IIA} and the dilaton~\eqref{IIA_D_dilaton} blow up on the NS5 locus~\eqref{brane_locus_np_2} with the appropriate power, as expected from Section~\ref{section_expansion}. 

The same divergence appears for the Type IIA $B_2$ field:
\begin{align}
B_{2,IIA}&=A_\sigma \, d\sigma \wedge d\beta+\widehat{B}_2 \nonumber \\
         &=\frac{r^2}{\tau^2\cos^2\alpha +r^2} \,  d\sigma \wedge d\beta +\widehat{B}_2 \label{B2_IIA}
\end{align}
where $\widehat{B}_2$ is as in \eqref{B2_NP}. The first term in~\eqref{B2_IIA} is the same as in~\eqref{IIa_b2} and  arises just from the geometry. As in Section~\eqref{section_expansion} the square of $H_3=d B_2 $ is of order one. 

The RR two-form field strength $F_2$ is:
\begin{equation}\label{F2_IIA}
F_2 = F_{3, \beta}
\end{equation}
where $F_{3,\beta}$ is as in~\eqref{f3b}. The square of this flux computed as in~\eqref{rule_expansion} using~\eqref{metric_IIA} is of order $\delta^2$.

According to Buscher's rules, the four-form field strength is given by:
\begin{equation}\label{F4_IIA}
F_4 =\widehat{F}_3 \wedge (d \beta +B_{a \beta} \, dy^a)+F_{5,\beta}
\end{equation}
where $\widehat{F}_3$, $B_{a\beta}\, dy^a$ and $F_{5,\beta}$ are defined in~\eqref{hatf3}, \eqref{bibeta} and~\eqref{F5_NP} respectively. Using the new metric~\eqref{metric_IIA} it turns out that $(\widehat{F}_3 \wedge d\beta)^2$ is of order $\delta^2$, while $(\widehat{F}_3 \wedge B_{a\beta}\, dy^a)^2$ is of order $\delta^4$ as well as $(F_{5, \beta})^2$. The lowest order component of $F_4$ in~\eqref{F4_IIA} is hence proportional to $M$, as one would expect for a four-form field-strength in the presence of smeared D4 branes.  

The Type IIA RR sector also contains a six-form and an eight-form field strengths, that can be computed from the hodge duals of~\eqref{F4_IIA} and~\eqref{F2_IIA} respectively, keeping in mind that the star operator is defined using~\eqref{metric_IIA}. We report here only the component of $dC_5$ with legs along the directions $0,1,2,3,\tau, \beta$, which is important for the purposes of the next section:
\begin{equation}\label{F6_IIA}
\left. dC_5  \right \rvert_{0123\tau \beta} = \frac{\varepsilon^{\frac{8}{3}}}{g_s^3M^2(\alpha^\prime)^2a_0^2}\frac{4}{3^{\frac{4}{3}}}\tau\, dt\wedge dx^1\wedge dx^2\wedge dx^3\wedge d\tau \wedge d\beta
\end{equation}
The square of~\eqref{F6_IIA} is of order $\delta^4$ and it is easy to verify that it comes from the hodge dual of the $F_{5,\beta}$ component in~\eqref{F4_IIA}, whose square is also of order $\delta^4$ as expected.


\subsection{Check: a D4 probe brane feels zero force}\label{section_probe_D4}
In this section we show that a D4 probe brane wrapping $t, x_1, x_2, x_3$ and $\beta$ feels no force in the Type IIA solution presented in Section~\ref{section_IIA_KS}. This result is expected from the fact that a probe D3 wrapping the first four coordinates in KS does not break any supersymmetries and is hence in equilibrium regardless of its location on the deformed conifold. Our D4 probe in the Type IIA dual KS solution interacts only with the metric~\eqref{IIA_D_metric}, the dilaton~\eqref{IIA_D_dilaton} and $C_5$ in~\eqref{F6_IIA}. Denoting pullbacks on the D4 worldvolume with a tilde the probe action is:
\begin{equation}\label{probe_action}
\mathcal{S}= -\int d^5 \,\tilde{x} \,e^{-\widetilde{\Phi}}\sqrt{-\textmd{det}\, \widetilde{g}_{\mu \nu}} + \int \widetilde{C}_5
\end{equation}
where the first integral is the Dirac - Born - Infeld action and the second one is the Wess-Zumino term. The potential $\widetilde{C}_5$ can be easily reconstructed by intrgrating~\eqref{F6_IIA}:
\begin{equation}\label{c5}
\widetilde{C}_5=\frac{\varepsilon^{\frac{8}{3}}}{g_s^3M^2(\alpha^\prime)^2a_0^2}\frac{2}{3^{\frac{4}{3}}}\tilde{\tau}^2\, d\tilde{t}\wedge d\tilde{x}^1\wedge d\tilde{x}^2\wedge d\tilde{x}^3\wedge d\tilde{\beta}
\end{equation}
In the Dirac-Born-Infeld part of the action~\eqref{probe_action} the dilaton~\eqref{IIA_D_dilaton} cancels $\tilde{g}_{\beta \beta}$ of~\eqref{IIA_D_metric} appearing in the determinant, so that the whole action is finite without divergences. The remaining of the integrand can be expanded in powers of $\delta$ up to highest corrections:
\begin{equation}\label{dbi_simplified}
  \sqrt{ \left(1-\frac{a_2\tilde{\tau}^2}{2 a_0}\right)^4} \simeq  \, 1 - \frac{a_2\tilde{\tau}^2}{ a_0}
\end{equation}
A quick check shows that~\eqref{dbi_simplified} and~\eqref{c5} are of the same order in $\delta$, so a cancellation in~\eqref{probe_action} is possible. Indeed, inserting~\eqref{dbi_simplified} and~\eqref{c5} into~\eqref{probe_action} and restoring all the constants from Section~\ref{section_IIA_KS} including $a_2$ in~\eqref{a2} one has:
\begin{equation}\label{action_simplified}
\mathcal{S} = -\int d^5 \tilde{x}\, \frac{\varepsilon^{\frac{8}{3}}}{g_s^3 M^2(\alpha^\prime)^2a_0 2^{\frac{2}{3}}} \left(1 + \frac{2^{\frac{2}{3}}\tilde{\tau}^2}{3^{\frac{4}{3}} a_0}\right) + \int d^5\tilde{x}\, \frac{\varepsilon^{\frac{8}{3}}}{g_s^3M^2(\alpha^\prime)^2a_0^2}\frac{2}{3^{\frac{4}{3}}} \tilde{\tau}^2 = \, const
\end{equation}
which shows that our D4 probe does not feel any force in Type IIA KS, as expeted. Moreover, the fact that the $\tau$-dependent part in~\eqref{dbi_simplified} exactly cancels against~\eqref{c5} proves that the criterion used to expand the KS field strengths in Section~\ref{section_expansion} is physically consistent with the expansion of the metric and its warp factors in~\eqref{metric_expanded}.


\section{Adding D3 branes to KS }\label{section_4}
In this section we want to push the T-duality procedure described in Section~\ref{section_practice} and Section~\ref{section_expansion} one step forward.
We modify the KS solution by adding $C$ D3 branes  at the NP wrapping the Minkowski space of KS. These will backreact interacting with the KS fields and causing a singularity at the NP, giving rise to what will be referred to as the KS+D3 solution. This solution is static because the $C$ D3 branes are perfectly stable at the NP, as shown for the T-dual solution in Section~\ref{section_probe_D4}. Applying the techniques described before we want to reconstruct the new T-dual version of the NP neighborhood. This operation is carried on to better understand the physics of the Type IIA dual solution to KS in view of testing the stability of the antibranes.

To perform the same procedure as in Section~\ref{section_program} it is necessary to check that~\eqref{isometry} remains an isometry after the backreaction of the $C$ D3 branes at the NP. This is fundamental to insure the existence of a new Type IIA solution with the same NS5 wrapping the holomorphic curve~\eqref{branes_deformed_conifold}. As proved in Section~\ref{section_practice}, the NP lies on the locus~\eqref{branes_deformed_conifold}, which is precisely the locus of fixed points on the deformed conifold under the isometry~\eqref{isometry}. Consequently, the backreaction of the D3 branes placed at the NP cannot spoil this isometry, which is hence preserved globally. A new Type IIA KS+D4 background T-dualized along $\xi$ in~\eqref{isometry} exists and hence it is perfectly legitimate to reconstruct the T-dual version of only a small region, namely the NP neighborhood.

As the D3 branes at the NP do not break any supersymmetry it is easy to find an ansatz to include their backreaction on the KS solution. The KS+D3 solution can be seen as some kind of superposition between the KS solution and the solution that one would get by placing the D3 branes at the NP on the empty deformed conifold. Indeed, the metric ansatz is still as in~\eqref{KS_metric}, but the warp factor now becomes
\begin{equation}\label{warpfactor_strategy}
h = h^{KS} + h^{D3}
\end{equation}
where $h^{KS}$ is the KS warp factor~\eqref{KS_warpfactor} and $h^{D3}$ is the warp factor that one would get by placing only the D3 branes on the empty deformed conifold. In addition, the five-form field strength becomes:
\begin{align}
F_5 &= \mathcal{F}_5+\star \mathcal{F}_5  &  \mathcal{F}_5 &= d \left(h^{-1}\right)\wedge dt\wedge dx^1\wedge dx^2\wedge dx^3 \label{f5_strategy}
\end{align}
where the hodge star should be computed using~\eqref{warpfactor_strategy}. The rest of the solution is constituted by the remaining fields in KS, namely $B_2$ in~\eqref{KS_B2} and $F_3$ in~\eqref{KS_F3}.

Finding $h^{D3}$ in~\eqref{warpfactor_strategy} is equivalent to solving the Laplace equation on the deformed conifold~\cite{Kuperstein_branes}. Now that the right ansatz for the KS+D3 solution has been found it is convenient to proceed to the next step, namely finding the expansion for $h^{D3}$ around the NP. The deformed conifold metric becomes as in~\eqref{metric_NP} which is an (almost) $\mathbb{R}^6$ metric. Therefore the lowest order expansion of $h^{D3}$ in the NP just becomes the blowing up warp factor that one gets by putting some D3 branes in flat empty space, and it will be of order $\delta^{-4}$, as confirmed by the analysis of~\cite{Kuperstein_branes}.  The next-to-lowest order corrections will start at least from order one and originates from the fact that we are expanding the solution of a Laplace equation on the deformed conifold, hence they can possibly be of the same order as the terms in $h^{KS}$ in~\eqref{warpfactor_NP}. We choose to ignore these higher order correction coming from the D3 brane backreaction. On one side, we know that these corrections take care of themselves and do not really add interesting physics to the problem as long as one captures the D3 brane divergence. On the other side, the interaction between these corrections and those of the same order coming from KS give rise to negligible terms, as we are primarily interested in the interaction between the KS terms and the new D3 divergence. Hence, our $h^{D3}$ is truncated at highest order, becoming:
\begin{equation}
h^{D3} = \varepsilon^{-\frac{8}{3}}\left(\frac{3}{2}\right)^{\frac{2}{3}}\frac{C}{R^4}
\end{equation}
 where we have defined $R= \sqrt{\tau^2+r^2+z^2}$ and we have taken into account the overall coefficient in~\eqref{metric_NP}.  The KS+D3 warp factor~\eqref{warpfactor_strategy} that we will consider is given by:
\begin{equation}\label{KS_C_warpfactor}
h = (g_s M \alpha^\prime)^2 \varepsilon^{-\frac{8}{3}}2^{\frac{2}{3}} (a_0 + a_2 \tau^2)+\varepsilon^{-\frac{8}{3}}\left(\frac{3}{2}\right)^{\frac{2}{3}}\frac{C}{R^4}
\end{equation}
The contribution coming from the D3 branes in~\eqref{KS_C_warpfactor} dominates over the KS ones, which can now be considered as corrections to the simple D3 brane solution. As in Section~\ref{section_expansion}, we have to expand the powers of the warp factor~\eqref{KS_C_warpfactor} differently so that all the coordinates contribute to the highest order in $\delta$ in the metric, taking into account that~\eqref{metric_NP} is of order $\delta^2$. We fix the highest order of expansion in the metric requiring it to comprise a $\tau$-dependent contribution from the KS in the warp factor~\eqref{KS_C_warpfactor}, which was essential for the stability of the D4 probes in Section~\ref{section_probe_D4}. This goal can be achieved if one expands $h^{\frac{1}{2}}$ and $h^{-\frac{1}{2}}$ as follows:
\begin{align}
\hat{h}^{\frac{1}{2}} & = \left(\frac{3}{2}\right)^{\frac{1}{3}}\varepsilon^{-\frac{4}{3}} \frac{\sqrt{C}}{R^2}+\frac{(g_s M \alpha^\prime)^2}{\varepsilon^{\frac{4}{3}}3^{\frac{1}{3}}}\,\frac{a_0  R^2}{\sqrt{C}}+\frac{(g_s M \alpha^\prime)^2}{\varepsilon^{\frac{4}{3}}3^{\frac{1}{3}}}\,\frac{a_2  \tau^2 R^2}{\sqrt{C}} \nonumber \\
\hat{h}^{-\frac{1}{2}}&=\left(\frac{2}{3}\right)^{\frac{1}{3}} \varepsilon^{\frac{4}{3}}\frac{R^2}{\sqrt{C}} -\frac{2^{\frac{2}{3}}(g_s M \alpha^\prime)^2\varepsilon^{\frac{4}{3}}}{3}\frac{a_0  R^6}{C^{3/2}} \label{warpfactor_expansion_D3}
\end{align}
Then the expanded metric of the KS+D3 solution becomes
\begin{equation}\label{metric_ansatz}
ds^2_{KS+D3} = \hat{h}^{-\frac{1}{2}}dx^i dx^i +\hat{h}^{\frac{1}{2}} ds^2_6
\end{equation} 
where $ds^2_6$ is as in~\eqref{metric_NP}. From~\eqref{warpfactor_expansion_D3} we get terms of order $\delta^2$ and $\delta^6$ from the Minkowski metric and terms of order $1$, $\delta^4$ and $\delta^6$ from the expansion of the deformed conifold metric. 

Now that we have expanded the KS+D3 metric we can proceed to the expansions of the field strengths. The criterion used for this purpose is the same as in Section~\ref{section_expansion}: one first expands the square of a field strength defined as in eq.~\eqref{rule_expansion} using~\eqref{metric_ansatz} to highest order and then truncates the expansion of the field strength itself keeping only the terms that contribute to the square. As the metric we are dealing with now is different from that in~\ref{metric_expanded}, the orders of magnitude of the squares change, as one would expect given the fact that the D3 branes at the NP tend to hide the KS solution in the NP neighborhood.

The expansion of $B_2$ is the same as in~\eqref{B2_NP}, but now both $(B_2)^2$ and $(H_3)^2$ are of order $\delta^6$. This is because the leading terms in the transverse metric are now of order one. Hence adding a leg to a form along a transverse direction does not change the order of magnitude of its square, while the differential of a form preserves the orders of magnitude of each term, as explained in Section~\ref{section_expansion}.

 Similarly to $B_2$, also $F_3$ gets rewritten as~\eqref{f3b} and~\eqref{hatf3}, and its square is of order $\delta^6$ as well. This is not the same for the expansion of $F_5$ in~\eqref{f5_strategy}, as $(F_5)^2$ is now of order one. This is as expected, as $F_5$ measures also the D3 brane charge on a sphere surrounding the NP, hence the flux of this form cannot be small. The component of $(F_5)^2$ that is of order one arises precisely from $h^{D3}$ in~\eqref{warpfactor_strategy} and if one truncated to this order the KS contribution to $F_5$ would be completely lost. To keep some reminiscence of the KS solution we expand $\mathcal{F}_5$ in~\eqref{f5_strategy} as follows:
\begin{align}
\mathcal{F}_5 &= d\left[ \left(\frac{2}{3}\right)^{\frac{2}{3}}\varepsilon^{\frac{8}{3}}\frac{R^4}{C}-\frac{4}{3^{\frac{4}{3}}}(g_sM \alpha^\prime)^2 \varepsilon^{\frac{8}{3}}a_0\frac{R^8}{C^2}\right]\wedge dt \wedge dx^1 \wedge dx^2\wedge dx^3 \nonumber  \\
 &=\mathcal{F}_5^{D3}+\mathcal{F}_5^{KS+D3} \label{NP_D3_f5}
\end{align}
where $\mathcal{F}_5^{D3}$ comes from the differential of the first term in the brackets and $\mathcal{F}_5^{D3+KS}$ comes from the second one. $\mathcal{F}_5^{D3}$ is purely due to the D3 branes, while $\mathcal{F}_5^{KS+D3}$ comes from the interaction between the KS solution and the branes. This can be qualitatively confirmed from the presence in $\mathcal{F}_5^{KS+D3}$ of some constants inherited from the KS solution, such as $M$ or $a_0$.  In addition, $(\mathcal{F}_5^{D3})^2$ is of order one, while the square of the second term is of order $\delta^8$. In section~\ref{section_expansion} $(F_5)^2$ in~\eqref{F5_NP} was of two orders higher than $(B_2)^2$ and $(F_3)^2$ and the same happens here for $(\mathcal{F}_5^{KS+D3})^2$ in~\eqref{NP_D3_f5}. Even if the orders of magnitude of the fluxes due to the KS solution have changed, the relative differences are preserved. 

To complete the expansion of $F_5$ we present the expression for $\star \mathcal{F}_5$:
\begin{align}
\star \mathcal{F}_5 =& \, \varepsilon^{-\frac{8}{3}}\left(\frac{3}{2}\right)^{\frac{2}{3}} \frac{\tau^2\cos^2 \alpha +2r^2}{h^4 \tau^4 \cos^3\alpha}(\partial_\tau h \,dr \wedge dz  - \partial_r h \, d\tau \wedge dz \nonumber \\
&- \partial_z h \, d\tau \wedge dr ) \wedge d\alpha \wedge d\beta \wedge d\sigma \label{star_f5}
\end{align}
where $h$ in~\eqref{warpfactor_strategy} should be properly truncated so to get terms of the same order as in~\eqref{NP_D3_f5}. For consistency, one should keep the two lowest order terms in the expansion of~\eqref{star_f5}, whose squares are of order one and $\delta^8$. 

The T-duality in $\beta$ of the new KS+D3 background is easily performed. The Type IIA metric close to the NP is given by
\begin{align}
ds^2_{IIA, KS+D3} = & \, \hat{h}^{-\frac{1}{2}}\left[dx^i dx^i + \varepsilon^{\frac{4}{3}}\left(\frac{2}{3}\right)^{\frac{1}{3}}\frac{d\beta^2}{(\tau^2 \cos^2\alpha + r^2)}\right] \nonumber \\
&+ \hat{h}^{-\frac{1}{2}} \varepsilon^{\frac{4}{3}}\left(\frac{2}{3}\right)^{\frac{1}{3}}\frac{B_{a \beta} \, dy^a }{(\tau^2 \cos^2 \alpha + r^2)}d\beta \nonumber \\
& + \hat{h}^{\frac{1}{2}} \varepsilon^{\frac{4}{3}}\left(\frac{2}{3}\right)^{\frac{1}{3}}\left[d\tau^2+\tau^2 d\alpha^2+dr^2+dz^2+\frac{r^2\tau^2\cos^2\alpha}{\tau^2\cos^2\alpha+r^2}d\sigma^2\right] \label{IIA_D_metric}
\end{align}
where the warp factors are expanded as in~\eqref{warpfactor_expansion_D3} and $B_{a\beta}\, dy^a$ is as in~\eqref{B2_NP}. As in Section~\ref{section_IIA_KS} the third line of the metric comes from the T-dualization of the deformed conifold, while the second line arises from the interaction between the KS $B_2$ and the geometry. The Type IIA dilaton now becomes:
\begin{equation}\label{IIA_D_dilaton}
e^{2\Phi}= \varepsilon^{\frac{4}{3}}\left(\frac{2}{3}\right)^{\frac{1}{3}} \frac{\hat{h}^{-\frac{1}{2}}}{ (\tau^2 \cos^2\alpha + r^2)}
\end{equation}
which is clearly of order one. The NS-NS two-form $B_2$ is exactly the same as in~\eqref{B2_IIA}:
\begin{equation}\label{IIA_D_B2}
B_{2,IIA} =\frac{r^2}{\tau^2\cos^2\alpha +r^2} d\sigma \wedge d\beta +\widehat{B}_2 
\end{equation}
with $\widehat{B}_2$ given by~\eqref{B2_NP}. The first term in~\eqref{IIA_D_B2} arises from $A_\sigma \, d\sigma \wedge d\beta$ where $A_\sigma \, d\sigma$ is defined in~\eqref{amu} and is a geometric feature of our T-duality in $\beta$ of the metric~\eqref{metric_NP}. The square of this term with~\eqref{IIA_D_metric} is still of order one and the same applies metric structure for its differential $H_3$\footnote{Notice that in~\eqref{IIA_D_metric} the highest order components of the metric along the NP coordinates are of order one, hence even if $H_3$ has one more leg than $B_2$ their squares are of the same order.}. This together with the fact that $B_2$, the dilaton and the metric blow up on~\eqref{brane_locus_np_2} indicate that this locus gets wrapped by NS5 branes even in the KS+D3 solution. 

The RR sector of the Type IIA version of the new KS+D3 background comprises a two-form $F_2$ which is exactly the same as in~\eqref{F2_IIA}, with $(F_2)^2 \sim \delta^6$. The four-form is now given by
\begin{equation}\label{IIA_D_F4}
F_4 = \widehat{F}_3\wedge (d\beta +B_{a\beta}\, dy^a)+F_{5,\beta}
\end{equation}
where $\widehat{F}_3$ and $B_{a \beta}\, dy^a$ are written in~\eqref{hatf3} and~\eqref{bibeta}, while $F_{5,\beta}$ should be computed from~\eqref{star_f5}. The lowest order contribution to $F_4$ is hidden in $F_{5,\beta}$ and its square is of order one: this represent the four-form field strength that one gets placing D4 branes in flat space. The next-to-lowest order contributions also come from $F_{5,\beta}$ and arise from the interactions between the KS solution and the D3 branes in Type IIB and their square is of order $\delta^8$.

Finally, the RR sector of this solution also includes a six-form field strength and an eight-form field strength, which can be computed via the hodge duals of~\eqref{IIA_D_F4} and~\eqref{F2_IIA} respectively. In particular, the component of $dC_5$ with legs along the Minkowski space and one among the $\tau$, $r$, $z$ coordinates on the conifold together with $\beta$ is easily computed from~\eqref{NP_D3_f5}:
\begin{equation} \label{IIA_D_F6}
\left. dC_5  \right \rvert_{0123\tau \beta} = ( \mathcal{F}_5^{D3}+\mathcal{F}_5^{KS+D3} ) \wedge d\beta
\end{equation}
The Type IIA KS+D3 solution incorporates all the main features of the previous T-duality of KS in Section~\ref{section_IIA_KS}, including the structure of the metric ~\eqref{IIA_D_metric} and the NS5 branes wrapping the same locus. New features arise from the novel terms signaling the interaction between the D4 branes and the T-dual KS solution. As a test, one could perform the probe computation of Section~\ref{section_probe_D4} using the metric~\eqref{IIA_D_metric} and the component of $dC_5$ in~\eqref{IIA_D_F6}. However, the ansatz we used in~\eqref{warpfactor_strategy} and~\eqref{f5_strategy} together with Buscher's rules guarantee that the D4 probe action is trivial. The cancellation in the D4 action~\eqref{probe_action} for the KS+D3 solution takes place at two different levels. Indeed, $\mathcal{F}_5^{D3}\wedge d\beta$ in~\eqref{IIA_D_F6} is canceled by the lowest order term in the DBI action arising from $h^{D3}$ in~\eqref{warpfactor_strategy}. These terms come from the pure D3 brane background in Type IIB and their cancellation in Type IIA just states that a D4 probe is in equilibrium in a D4 brane background. Then, the next-to-leading order correction $\mathcal{F}_5^{KS+D3}\wedge d\beta$ in~\eqref{IIA_D_F6} is cancelled against the next-to-leading order term in the DBI action coming from $h^{KS}$ in~\eqref{warpfactor_strategy}. This cancellation is physically more meaningful than the previous one, as it is due to terms in Type IIA  arising from the interaction between the D3 localized branes and the KS solution.


\section{Conclusions and outlook}\label{section_discussion}
We reconstructed the Type IIA solution T-dual to the KS solution on a small region at the tip of the deformed conifold, choosing the correct isometry to obtain an NS5 brane wrapping a holomorphic curve in Type IIA. We discussed the choice of our isometry both from the point of view of the dual cascading four-dimensional gauge theory and from the geometric properties of the deformed conifold. This operation was made possible by finding a suitable set of coordinates for the North Pole expansion. In Section~\ref{section_4} the same techniques were applied to T-dualize the solution constructed by adding D3 branes at the North Pole of the three-sphere at the bottom of the deformed conifold. On one hand, the North Pole expansion makes it easy to solve the Laplace equation for the D3's on the deformed conifold, as the leading order term in the expanded metric corresponds to the solution to the Laplace equation for D3 branes in flat space. On the other hand, the expansion makes it easy to identify the physics arising purely from the D4 branes in Type IIA and that coming from the interactions between the Type IIA T-dual solution to ``empty" KS and the additional localized D4 branes. 

The solutions dual to KS and KS+D3 constructed in this paper mark a first step towards testing the stability of antibranes in Type IIA. Adding an anti-D4 brane in the T-dual solution to KS of Section~\ref{section_IIA_KS} is the next step in this direction. It is difficult to find the full backreaction of the anti-D4 on the T-dual KS solution because of the supersymmetry breaking. However, we expect that the form of the T-dual solution to the KS+D3 one of Section~\ref{section_4} could be used to get a better understanding about the backreaction of the anti-D4 and possibly to propose an ansatz. For instance, the backreaction of the anti-D4 should preserve the relative difference between the order of magnitudes of the squares of the fields arising from KS, as happens for a backreacted D4. In addition, the anti-D4 will not alter the divergencies of the dilaton, metric and $B_2$ near the Type IIA NS5 brane: as we have seen in Section~\ref{section_4}, the squares of these divergent terms have the same order of magnitude as in the T-dual solution of KS presented in Section~\ref{section_IIA_KS}. 

Another interesting possibility is to study the brane-antibrane interactions between the backreacted D4 branes of the solution in Section~\ref{section_4} and a probe anti-D4 brane. Clearly, the leading terms of the probe action would represent the attractive force exerted by the backreacted branes. The interesting physics would then be hidden in the subleading terms of this action. This would be the Type IIA correspondent of the interaction between the fields of the KS solution and the  backreacted D3 branes in Type IIB. If the force exerted by the next-to-leading order terms were repulsive this would prove that anti-D4 branes at the bottom of the solution T-dual to KS are unstable. In particular, we expect the fields sourced by the NS5 brane in Type IIA to play a key role in the final results.

\section*{Acknowledgments} 
We are grateful to Iosif Bena for useful discussions and comments on the manuscript. This work was supported in part by the ERC Starting Grant 240210 \textit{String}-QCD-BH and by the John Templeton Foundation Grant 48222.


\appendix

\section{Review of the deformed conifold} \label{appendix_coordinates}
In this section we briefly review the main features of the deformed conifold and explain how the coset and brane coordinates of Section~\ref{section_KS} and~\ref{section_program} are introduced to parameterize this manifold.

The deformed conifold is a hypersurface in $\mathbb{C}^4$:
\begin{equation}\label{conifold_hypersurface}
w_1^2+w_2^2+w_3^2+w_4^2=\varepsilon^2
\end{equation}
where $w_i\in \mathbb{C}$ and we assume $\varepsilon \in \mathbb{R}_{>0}$ with no loss of generality. We call the $w_i \in \mathbb{C}$ subject to the constraint~\eqref{conifold_hypersurface} the \textit{conifold coordinates}. \\
The deformed conifold is a cone  over the $T^{1,1}$ base space. The latter is topologically $S_2\times~S_3$, where only the $S_2$ shrinks at the tip of the cone so that deformed conifold has no singularities. These observations are easily proved in conifold coordinates. To study the base space of the deformed conifold one intersects~\eqref{conifold_hypersurface} with a sphere of radius $r$ in $\mathbb{C}^4$ defined as
\begin{equation}\label{foursphere}
|w_1|^2+|w_2|^2+|w_3|^2+|w_4|^2=r^2.
\end{equation}
Writing each $w_i$ as $w_i=a_i +ib_i$ one then gets the three following conditions that define the $T^{1,1}$ space in conifold coordinates:
\begin{align}
&\sum_{i=1}^4 a_i^2=\frac{r^2+\varepsilon^2}{2} \label{threesphere}\\
&\sum_{i=1}^4 b_i^2=\frac{r^2-\varepsilon^2}{2}\label{twosphere} \\
&\sum_{i=1}^4 a_i \cdotp b_i =0 \label{bundlecondition}
\end{align}
which also require $r^2\geq \varepsilon^2$. Equation~\eqref{threesphere} defines a three sphere $S_3$ that remains finite for $r=\varepsilon$, i.e. at the tip of the deformed confold. Equation~\eqref{twosphere} describes a two sphere $S_2$ fibered over the three sphere, where the fibration is specified by~\eqref{bundlecondition}. Notice that the $S_2$ shrinks at the tip of the conifold. In~\cite{Candelas} it was proved that $T^{1,1}=S_3 \times S_2$, namely that the fibration is trivial.

The brane coordinates introduced in Section~\ref{section_program} are crucial to find the isometry to T-dualize the KS solution and to write the locus wrapped by the NS5 in Type IIA. They are easily related to the conifold ones. Define a matrix $\mathcal{W}$ as
\begin{equation}
\mathcal{W}=\frac{1}{\sqrt{2}}w_i \, \sigma^i
\end{equation} 
where $\sigma^i$ for $i=1,2,3$ are the usual Pauli matrices and $\sigma^4 \equiv i \mathbf{1}$. The brane coordinates $(x,u,z_1,z_2)\in \mathbb{C}^4$ are defined by the entries of $\mathcal{W}$:
\begin{equation}\label{branecoordinates}
\mathcal{W}=
\begin{pmatrix}
z_1 && x \\
u && z_2
\end{pmatrix} = \frac{1}{\sqrt{2}}
\begin{pmatrix}
w_3+iw_4 && w_1-iw_2 \\
w_1+iw_2 && -w_3+iw_4
\end{pmatrix}
\end{equation}
The definition of the deformed conifold~\eqref{conifold_hypersurface} and the sphere in $\mathbb{C}^4$~\eqref{foursphere} respectively become:
\begin{align}
\textmd{det} \, \mathcal{W}=-\frac{\varepsilon^2}{2} \quad &\Rightarrow \quad z_1z_2-xu=-\frac{\varepsilon^2}{2} \label{conifold_fibration}\\
\textmd{Tr}(\mathcal{W}^{\dagger}\mathcal{W})=r^2 \quad &\Rightarrow \quad |x|^2+|u|^2+|z_1|^2+|z_2|^2=r^2 \label{foursphere_brane}
\end{align}

Finally, we show how the coset coordinates of Section~\ref{section_KS} are introduced and relate them to the brane ones. This last set of coordinates is the one that is always used for computational purposes. In~\cite{Candelas} the $T^{1,1}$  base space for the deformed conifold is defined as a coset manifold: 
\begin{equation}\label{coset}
T^{1,1}=\frac{SU(2)\times SU(2)}{U(1)}
\end{equation}
Parameterizing each $SU(2)$ via Euler angles $(\phi_1, \theta_1, \psi_1)$ and $(\phi_2, \theta_2, \psi_2)$ as in~\cite{Minasian} one can write a generic element in the coset as
\begin{equation}\label{coset_representative}
e^{\frac{i}{2} \sigma_1 \phi_1}e^{\frac{i}{2} \sigma_2 \theta_1}e^{\frac{i}{2}\sigma_1^{\prime} \phi_2}e^{\frac{i}{2} \sigma_2^{\prime} \phi_2}e^{\frac{i}{2} (\sigma_3+\sigma^{\prime}_3)\psi}
\end{equation}
where $\psi \equiv \psi_1+\psi_2$ and $\sigma_i$ $\sigma^{\prime}_i$ are two sets of Pauli matrices such that $[\sigma_i, \sigma^{\prime}_j]=0$. Given the coset parameterization~\eqref{coset_representative} an element of the $U(1)$ quotient group is hence written as $ e^{\frac{i}{2}(\sigma_3+\sigma_3^{\prime})(\psi_1-\psi_2)}$.  The radial coordinate $\tau$ is introduced via
\begin{equation}\label{taudefinition}
r^2=\varepsilon^2 \cosh \tau
\end{equation}
where $r$ is defined in~\eqref{foursphere}, and hence the tip of the conifold is defined by $\tau =0$. The coset coordinates allow to find the Ricci-flat K\"ahler metric in~\eqref{df_metric}.  For our purposes it is most useful to rewrite the brane coordinates as functions of the coset ones~\cite{royston_deformed}:
\begin{align}
x&=\frac{\varepsilon}{\sqrt{2}}\left(\cos\frac{\theta_1}{2}\cos\frac{\theta_2}{2}e^{\frac{1}{2}(\tau+i\psi)}-\sin\frac{\theta_1}{2}\sin\frac{\theta_2}{2}e^{-\frac{1}{2}(\tau+i\psi)}\right) e^{\frac{i}{2}(\phi_1+\phi_2)} \nonumber \\
u&=\frac{\varepsilon}{\sqrt{2}}\left(-\sin\frac{\theta_1}{2}\sin\frac{\theta_2}{2}e^{\frac{1}{2}(\tau+i\psi)}+\cos\frac{\theta_1}{2}\cos\frac{\theta_2}{2}e^{-\frac{1}{2}(\tau+i\psi)}\right) e^{-\frac{i}{2}(\phi_1+\phi_2)} \nonumber \\
z_1&=-\frac{\varepsilon}{\sqrt{2}}\left(\cos\frac{\theta_1}{2}\sin\frac{\theta_2}{2}e^{\frac{1}{2}(\tau+i\psi)}+\sin\frac{\theta_1}{2}\cos\frac{\theta_2}{2}e^{-\frac{1}{2}(\tau+i\psi)}\right) e^{\frac{i}{2}(\phi_1-\phi_2)} \nonumber \\
z_2&=\frac{\varepsilon}{\sqrt{2}}\left(\sin\frac{\theta_1}{2}\cos\frac{\theta_2}{2}e^{\frac{1}{2}(\tau+i\psi)}+\cos\frac{\theta_1}{2}\sin\frac{\theta_2}{2}e^{-\frac{1}{2}(\tau+i\psi)}\right) e^{\frac{i}{2}(-\phi_1+\phi_2)} \label{branes_to_coset}
\end{align}


\section{The tip of the deformed conifold} \label{appendix_tip}
In this section we provide a suitable parameterization for the tip of the conifold, following~\cite{royston_deformed}.  The tip of the conifold is the $\tau = 0$ locus in coset coordinates, which corresponds to $r^2=\varepsilon^2$ in~\eqref{taudefinition}. The metric~\eqref{df_metric} is finite as
$K(\tau)\rightarrow \left(\frac{2}{3}\right)^{\frac{1}{3}}$ and  the metric at the tip becomes:
\begin{equation}\label{threespheremetric}
d\Omega_3^2=\frac{\varepsilon^{\frac{4}{3}}}{2}\left(\frac{2}{3}\right)^{\frac{1}{3}}\left[\frac{1}{2}(g^5)^2+(g^3)^2+(g^4)^2\right]
\end{equation}
As expected from~\eqref{threesphere} this should be the round metric of the surviving $S_3$. This is can be proved defining as in~\cite{Minasian}
\begin{equation}\label{matrixT}
T=L_1\sigma_1L_2^{\dagger}\sigma_1
\end{equation}
where $L_1$ and $L_2$ are matrices of the $SU(2)$ groups in~\eqref{coset} parametrized via Euler angles as in~\eqref{coset_representative}. One then has 
\begin{equation}\label{threespherestandard}
\textmd{Tr}(dT^{\dagger}dT)=\frac{1}{2}(g^5)^2+(g^3)^2+(g^4)^2
\end{equation}
and as $T$ itself is an $SU(2)$ matrix the metric above represents the standard three-sphere metric. Then from~\eqref{threespheremetric} one reads that the squared radius of the $S_3$ at the tip is proportional to $\varepsilon^{\frac{4}{3}}$.

Note that the metric of the deformed conifold~\eqref{df_metric} is invariant under the $\mathbb{Z}_2$ symmetry  that exchanges $\phi_1,\theta_1$ and $\phi_2, \theta_2$. Indeed, the coset coordinates depict the $T^{1,1}$ base as a symmetric $S_1$ fibration over $S_2\times S_2$, where the fiber is parametrized by $\psi$, while $(\phi_i,\theta_i)$ parametrize the two $S_2$. The coset coordinates are not suitable to describe this $S_3$. The matrix $T$ introduced in~\eqref{matrixT} parameterizes precisely the $SU(2)$ to which the $T^{1,1}$ base degenerates at the tip of the conifold, which is symmetrically embedded in the coset~\eqref{coset}.  We then introduce the Euler angles $\zeta \in [0, \pi[$ and $\phi_w, \phi_x \in [0, 2\pi[$ to rewrite $T$ as:
\begin{align}
T=
\begin{pmatrix}
\cos \frac{\zeta}{2}e^{i \phi_x} & -\sin \frac{\zeta}{2}e^{-i \phi_w}\\
\sin \frac{\zeta}{2}e^{i \phi_w} & \cos \frac{\zeta}{2}e^{-i \phi_x}
\end{pmatrix}
\end{align}
and comparing with~\eqref{matrixT} one gets
\begin{align}
\cos^2\frac{\zeta}{2} &=\frac{1}{2}[1+\cos\theta_1\cos\theta_2-\cos\psi\sin\theta_1\sin\theta_2] \nonumber \\
\phi_w &=\arctan\left[\frac{\sin\left(\frac{\theta_1-\theta_2}{2}\right)}{\sin\left(\frac{\theta_1+\theta_2}{2}\right)} \tan\frac{\psi}{2}\right]-\frac{1}{2}(\phi_1-\phi_2) \nonumber \\
\phi_x &=\arctan\left[\frac{\cos\left(\frac{\theta_1-\theta_2}{2}\right)}{\cos\left(\frac{\theta_1+\theta_2}{2}\right)} \tan\frac{\psi}{2}\right]+\frac{1}{2}(\phi_1+\phi_2) \label{threespherecoordinates}
\end{align}
while the metric~\eqref{threespherestandard} is given by
\begin{equation}\label{threesphere_metric}
d\Omega_3^2=\frac{d\zeta^2}{2}+2\sin^2\frac{\zeta}{2}d\phi_w^2+2\cos^2\frac{\zeta}{2}d\phi_x^2
\end{equation}
The coordinates $(\zeta, \phi_w, \phi_x)$ see the three sphere as a circle fibration over a disc, where the fiber is parameterized by $\phi_x$ and the base is parameterized by $(\zeta, \phi_w)$. The fiber smoothly shrinks at the boundary of the disc so to give a smooth $S_3$.

The parameterization of the three-sphere~\eqref{threespherecoordinates} is useful also to justify the redefinition~\eqref{spherical_cylindrical} that completes the NP expansion. Indeed plugging~\eqref{np_expansion} and~\eqref{spherical_cylindrical} into~\eqref{threespherecoordinates} and expanding to lowest order in $\delta$ one gets
\begin{align}
\cos^2\frac{\zeta}{2} &\simeq r^2  &
\phi_w &\simeq \frac{\pi}{2} +z &
\phi_x & = \sigma+\beta \label{threespherecoordinates_NP}
\end{align}
From~\eqref{threespherecoordinates_NP} it is clear that the NP ($\tau=z=r=0$) lies on the boundary of the base disc $\xi=\pi$ of the fibration~\eqref{threesphere_metric}. Inserting~\eqref{spherical_cylindrical} into~\eqref{threesphere_metric} and~\eqref{threespherecoordinates_NP} one obtains the linearized flat metric in cylindrical coordinates of~\eqref{metric_NP}.

The coordinates~\eqref{threespherecoordinates} allows to nicely parameterize the NS5 locus~\eqref{branes_deformed_conifold} at the tip of the conifold. For $\tau=0$ the coordinate $\phi_x$ in~\eqref{threespherecoordinates} is the phase of the brane coordinate $x$ in~\eqref{branes_to_coset} when rewritten as $x=|x|e^{i\phi_x}$. This means that it can be used to parametrize the isometry~\eqref{isometry}, i.e. it can be used as T-duality coordinate. Indeed, the coordinate $\beta$ that we used for the T-duality around the NP basically coincides with $\phi_x$ plus a shift -see~\eqref{threespherecoordinates_NP}. Using~\eqref{branes_to_coset} it is possible to rewrite the NS5 locus~\eqref{branes_deformed_conifold} in coset coordinates:
\begin{equation}\label{degeneration_coset}
2+2\cos\theta_1\cos\theta_2-e^{-\tau+i\psi}(1+e^{2\tau+2i\psi})\sin\theta_1\sin \theta_2=0
\end{equation}
If one imposes $\tau=0$ in~\eqref{degeneration_coset} and then uses~\eqref{threespherecoordinates} one gets
\begin{equation}
\cos^2\frac{\zeta}{2}=0
\end{equation}
which means that the NS5 for $\tau=0$ wraps the boundary of the disc $\zeta=\pi$ in Type IIA. Indeed, the component of the metric~\eqref{threesphere_metric} for $\phi_x$ degenerates exactly on this locus in Type IIB. Note that plugging~\eqref{np_expansion} and~\eqref{spherical_cylindrical} in~\eqref{degeneration_coset} and expanding to lowest order one gets precisely~\eqref{brane_locus_np_2}.


\section{The isometry for the T-duality of KS}\label{appendix_isometry_proof}
In this section we prove that the transformation
\begin{align}
x&\rightarrow e^{i\xi} x &     u&\rightarrow e^{-i \xi}u \label{isometry2}
\end{align}
performed on the brane coordinates~\eqref{branecoordinates} is an isometry for the KS solution of Section~\ref{section_KS}, assuming that it is an isometry for the metric on the deformed conifold~\eqref{df_metric}, which was shown in~\cite{royston_deformed}.

To prove that~\eqref{isometry2} is an isometry for the full KS background one must show that it leaves invariant all the other fields together with the warped metric. Observing that in the KS solution $F_5=B_2\wedge F_3$ and that $F_3$ and $H_3=dB_2$ are related by the supersymmetry equations, one concludes that~\eqref{isometry2} is an isometry for the full KS solution if and only if it leaves $B_2$ in~\eqref{KS_B2} and the warp factor $h(\tau)$ in~\eqref{KS_warpfactor} invariant.

From~\eqref{branecoordinates}  only the conifold coordinates $w_1$ and $w_2$ depend on $x$ and $u$ and under~\eqref{isometry2} these transform as
\begin{align}
w_1 = \frac{x+u}{\sqrt{2}} \quad &\longrightarrow  \quad \frac{e^{i\xi}x+e^{-i\xi}u}{\sqrt{2}} \nonumber \\
 w_2 =-i\frac{u-x}{\sqrt{2}}  \quad &\longrightarrow  \quad -i\frac{e^{-i\xi} u+e^{i\xi}x}{\sqrt{2}}\label{isometryconifold}
\end{align}
this is equivalent to
\begin{align}
\begin{pmatrix}
 w_1 \\
 w_2
\end{pmatrix}
\longrightarrow
\begin{pmatrix}
\cos \xi && \sin \xi \\
-\sin \xi && \cos \xi
\end{pmatrix}
\begin{pmatrix}
 w_1 \\
 w_2
\end{pmatrix}
\end{align}
This proves that $w_1, w_2$ are rotated by an angle $\xi$ under~\eqref{isometry2} and hence this transformation belongs to the $SO(4)$ group that leaves the conifold invariant, as is clear from~\eqref{conifold_hypersurface}. 

Working in conifold coordinates it is then easy to see from~\eqref{taudefinition} and~\eqref{foursphere} that $\tau$ is invariant under~\eqref{isometry2}. Consequently, all the functions in the KS solution~\eqref{KS_functions} and the warp factor~\eqref{KS_warpfactor} are invariant under this transformation.  In addition, as shown in~\cite{remarks}, $B_2$ in~\eqref{KS_B2} can be rewritten in conifold coordinates in an $SO(4)$-invariant form:
 \begin{align}
 B_2&=g(\tau) \epsilon_{ijkl} w^i \bar{w}^{\bar{j}} dw^k\wedge d\bar{w}^{\bar{l}} & g(\tau)&=\frac{ig_s M \alpha^{\prime}}{3\varepsilon^4}\frac{\tau \coth \tau -1}{\sinh^2\tau} \label{b2conifold}
 \end{align}
 and then a fortiori $B_2$ is invariant under~\eqref{isometry2}. In the formula above $i,j,k =1,2,3,4$ and $\tau$ is implicitly rewritten as a function of the $w_i$. This completes the proof that~\eqref{isometry2} is an isometry for the whole KS solution and as this is a $U(1)$ transformation it can be used to T-dualize this solution.


\section{Some expansions in the NP neighborhood}\label{computation_expansion}
We report in this section some necessary computations to expand the KS solution in the neighborhood of the NP. After rewriting the deformed conifold metric around the NP in~\eqref{metric_NP}, it is necessary to expand the KS warp factor, defined as:
\begin{align}
h(\tau)&=(g_s M \alpha^\prime)^2 \varepsilon^{-\frac{8}{3}}2^{\frac{2}{3}} I(\tau)  &  I(\tau)&=\int_{\tau}^{\infty}dx \, \frac{x\coth x -1}{\sinh^2 x}(\sinh 2x -2x)^{\frac{1}{3}}\label{KS_warpfactor2}
\end{align}
The function $I(\tau)$ is even and close to the NP for $\tau$ of order $\delta$ it behaves as\footnote{We are using here $\tilde{\tau}$ of~\eqref{np_expansion}, dropping the twiddle and taking care of the factor of two}:
\begin{equation}
I(\tau)= a_0 + a_2 \tau^2+\mathcal{O}(\tau^4)
\end{equation}
with $a_0\approx 0.71805$ was computed in~\cite{Klebanov-Strassler}. To compute $a_2$ one expands:
\begin{equation}
a_2 \tau^2 \simeq I(\tau)-I(0) \simeq -\int_0^{\tau} \frac{2^{\frac{2}{3}}}{3^{\frac{4}{3}}}x \, dx
\end{equation}
where the integrand of~\eqref{KS_warpfactor2} has been expanded for $x$ small. One then easily gets:
\begin{equation}\label{a2}
a_2=-2 \left(\frac{2}{9}\right)^{\frac{2}{3}}
\end{equation}
The expansion of the other functions of $\tau$ in~\eqref{KS_functions} is much easier\footnote{Here as before we are expanding substituting $\tau = 2\tilde{\tau}$ as prescribed in~\eqref{np_expansion} and then we remove the twiddle}:
\begin{align}
f(\tau) &\simeq \frac{2}{3} \tau ^3  &  k(\tau)&\simeq \frac{2}{3} \tau  \nonumber \\
F(\tau) &\simeq \frac{\tau^2}{3}  & \ell(\tau) &\simeq \frac{8}{9}\tau^3 \nonumber \\ 
K(\tau) & \simeq \left(\frac{2}{3}\right)^{\frac{1}{3}} & \frac{\ell(\tau)}{K^2\sinh^2\tau} &\simeq \left(\frac{2}{3}\right)^{\frac{1}{3}}\frac{\tau}{3}\label{expanded_functions}                 
\end{align}
The next step is to expand the base one-forms of the deformed conifold~\eqref{KS_conifold_one_forms_1} around the NP using~\eqref{np_expansion}. The expansion is carried on up to order $\delta$:
\begin{align}
g^1/\sqrt{2}&\simeq -\cos \alpha \, d\beta +r\frac{\cos \sigma}{\cos \alpha}d\alpha  \nonumber \\
g^2/\sqrt{2}&\simeq d\alpha + r\cos \sigma \, d\beta \nonumber \\
g^3/\sqrt{2}&\simeq \cos \alpha \, dz +\sin \alpha \,[\cos \sigma \, dr-r\sin \sigma (d\beta +d\sigma)] \nonumber \\
g^4/\sqrt{2}&\simeq -r\cos\sigma \, (d\beta+d\sigma)-\sin \sigma \, dr \nonumber \\
g^5&\simeq 2 \sin \alpha \, dz-2\cos \alpha \,[\cos \sigma \, dr -r\sin \sigma (d\beta+ d\sigma)] \label{expanded_oneforms}
\end{align}
Note that only $g^1$ and $g^2$ are of order one in the $\delta$-expansion, while all the other forms are of order $\delta$. Indeed, these two-forms are defined on the angles of the sphere (and cylinder) in the coordinate system of~\eqref{np_expansion}. 

Finally, to expand the RR and NS-NS fields around the NP one needs to expand the wedge products of the base one-forms on the conifold. We report here some nontrivial ones, that can be derived using~\eqref{expanded_oneforms}:
\begin{align}
g^1\wedge g^3 &\simeq 2\cos\alpha \, (\cos \alpha\, dz+\sin\alpha \cos\sigma \, dr-r\sin\alpha\sin\sigma d\sigma) \wedge d\beta \nonumber \\
g^2\wedge g^4 &\simeq -2[r\cos\sigma (d\beta +d\sigma)+\sin\sigma\, dr]\wedge d\alpha \nonumber \\
g^5\wedge g^3\wedge g^4 &\simeq 4 r\, [dr\wedge dz\wedge (d\beta+d\sigma)]\nonumber \\
g^5\wedge g^1\wedge g^2 &\simeq 4\cos \alpha\,(\sin\alpha \, dz -\cos\alpha \cos\sigma \, dr-r\cos\alpha \sin\sigma d\sigma)\wedge d\alpha \wedge d\beta \label{expanded_wedges}
\end{align}
The sign of a wedge product depends on the orientation chosen for the coordinates. Here and in every NP expansion we have always used the following ordering: $\tau, \alpha, \beta, r, z, \sigma$.


\section{Buscher's rules for T-duality}\label{section_buscher}
In this section we recall Buscher's rules used to construct the T-dual KS solution in Type IIA around the north pole. To T-dualize from type IIB to Type IIA and vice versa one needs an isometry along a compact direction $y$. Before performing the T-duality it is convenient to rewrite the fields as follows
\begin{align}
ds^2 &= g_{yy}(dy + A_{i} dx^{i})^2 + \widehat{g}_{ij}dx^i dx^j \nonumber
\\
B_2 &= B_{i y} dx^{i} \wedge (dy+ A_{i} dx^{i})+\widehat{B}_2 \nonumber \\
C_p &= C^y_{p-1}\wedge (dy +B_{i y} dx^i)+\widehat{C}_p
\end{align}
The T-dual solution is then given by
\begin{align}
d\widetilde{s}^2 &= g_{yy}^{-1}(dy + B_{iy} dx^i)^2+\widehat{g}_{ij}dx^i dx^j\nonumber \\
e^{2\tilde{\Phi}} &= g_{yy}^{-1}e^{2\Phi} \nonumber \\
\widetilde{B}_2 &= A_{i} dx^i \wedge dy + \widehat{B}_2 \nonumber \\
\widetilde{C}_s &= \widehat{C}_{s-1} \wedge (dy + B_{iy} dx^i) + \widehat{C}_s^y
\end{align}
If the RR potentials are not know it is possible to perform the T-duality directly on the field strengths. These should first be rewritten as:
\begin{equation}
F_p = F^y_{p-1}\wedge (dy + A_i dx^i) +\widehat{F}_p
\end{equation}
and then transformed into:
\begin{equation}
\widetilde{F}_s = \widehat{F}^y_{s-1}\wedge (dy + B_{iy} dx^i)+F^y_s
\end{equation}


\renewcommand{\leftmark}{\MakeUppercase{Bibliography}}
\phantomsection
\bibliographystyle{JHEP}

\bibliography{References_giulio}

\label{biblio}

\end{document}